\documentclass[12pt]{article}

\usepackage[a4paper,scale=0.8]{geometry}
\usepackage{graphicx}
\usepackage{setspace}
\usepackage{amsmath,amssymb,amsthm,bm}
\usepackage{mathtools}
\usepackage{enumitem}

\usepackage{booktabs}
\usepackage{array}
\usepackage{float}

\usepackage{tikz}
\usetikzlibrary{positioning,calc,fit,shapes.geometric,scopes,decorations.pathreplacing,arrows.meta}
\usepackage{pgfplots}
\pgfplotsset{compat=1.18}
\usepackage{tikz,pgfplots}
\usetikzlibrary{intersections}
\usepgfplotslibrary{fillbetween}

\usepackage{xcolor}

\usepackage{natbib}
\setcitestyle{authoryear,open={(},close={)}}

\usepackage{caption}
\usepackage{subcaption}

\usepackage{changes}
\usepackage[colorlinks=true,linkcolor=blue,urlcolor=blue,citecolor=blue]{hyperref}
\usepackage[nameinlink,noabbrev]{cleveref}

\crefname{assumption}{Assumption}{Assumptions}
\Crefname{assumption}{Assumption}{Assumptions}

\crefname{theorem}{Theorem}{Theorems}
\Crefname{theorem}{Theorem}{Theorems}

\crefname{lemma}{Lemma}{Lemmas}
\Crefname{lemma}{Lemma}{Lemmas}

\crefname{proposition}{Proposition}{Propositions}
\Crefname{proposition}{Proposition}{Propositions}

\crefname{corollary}{Corollary}{Corollaries}
\Crefname{corollary}{Corollary}{Corollaries}

\crefname{remark}{Remark}{Remarks}
\Crefname{remark}{Remark}{Remarks}

\crefname{definition}{Definition}{Definitions}
\Crefname{definition}{Definition}{Definitions}

\crefname{observation}{Observation}{Observations}
\Crefname{observation}{Observation}{Observations}

\crefname{subassumption}{Assumption}{Assumptions}
\Crefname{subassumption}{Assumption}{Assumptions}

\DeclareMathOperator*{\argmax}{arg\,max}
\DeclareMathOperator{\conc}{conc}
\DeclareMathOperator{\Sel}{Sel}

\DeclareMathOperator{\F}{\mathcal{F}}
\DeclareMathOperator{\Y}{\mathcal{Y}}
\DeclareMathOperator{\G}{\mathcal{G}}
\DeclareMathOperator{\T}{\mathbb{T}}

\providecommand{\conc}{\operatorname{conc}}

\newcommand{\ind}{\mathbf 1} 

\numberwithin{equation}{section}

\theoremstyle{plain}
\newtheorem{theorem}{Theorem}
\newtheorem{lemma}{Lemma}
\newtheorem{proposition}{Proposition}
\newtheorem{corollary}{Corollary}

\theoremstyle{definition}
\newtheorem{definition}{Definition}
\newtheorem{assumption}{Assumption}

\theoremstyle{remark}
\newtheorem{remark}{Remark}[section]

\newcounter{subassumption}[assumption]

\newtheorem{assumptionsub}[subassumption]{Assumption}

\crefname{assumptionsub}{Assumption}{Assumptions}
\Crefname{assumptionsub}{Assumption}{Assumptions}

\title{Dynamic Mechanism Collapse: A Belief-Space Representation}

\author{%
Xiaopeng Zeng\textsuperscript{1}
\quad
Erbao Cao\textsuperscript{1,*}\\[0.4em]
\textsuperscript{1}\small School of Economics and Trade, Hunan University, Changsha, China\\
\textsuperscript{*}\small Corresponding author. Email: \texttt{ceb9491@126.com}
}

\date{June 5, 2026}

\begin{document}
\maketitle

\begin{abstract}
Public histories may change continuation rules but not ex-ante value. We study when public histories beyond posterior beliefs affect the seller's ex-ante value in dynamic mechanism design with a single-shot resource. We construct a belief-space payoff map from maximal sequential-equilibrium continuation values indexed by date and posterior belief. We show that the optimal ex-ante value of the dynamic problem is the finite concavification of the upper envelope of these continuation values. We give a belief-space criterion for when non-posterior public histories have no additional value, so that a terminal posterior-only mechanism attains the full-history value. Our results provide structural insights into when public histories matter for ex-ante value rather than merely for continuation play.
\end{abstract}



\newpage
\section{Introduction}
Public histories in dynamic mechanism design enter the seller's problem through two channels. They update beliefs about the payoff-relevant state, and they may also alter the continuation opportunities available to the seller through eligibility, reputational, or other public variables that carry no further information about the state. A central concern is when the second, non-posterior channel affects the seller's ex-ante value, and when it does not.

One could approach the question at the level of the optimal continuation rule, asking whether a posterior-measurable selector implements the full-history optimum after every public history. This selector-level requirement, however, is strictly stronger than ex-ante revenue equivalence. We pose the same question at a weaker level: we ask which public histories can be dropped without changing the seller's ex-ante value, while leaving the optimal continuation rule free to depend on them.

To fix ideas, consider a patient with a settled diagnosis seen by the same physician at two hospitals, differing only in whether the standard treatment is covered by insurance. If the condition is routinely treatable, the change in coverage shifts both the prescribed treatment and the realized outcome. If the condition is terminal, the prescribed treatment still changes---aggressive when covered, palliative when not---but the prognosis does not. The non-posterior public state moves the continuation rule in both cases; but it moves the principal's payoff only in the first.

This observation organizes the paper. We develop a belief-space representation of the seller's ex-ante dynamic mechanism optimum, give a sharp condition for when non-posterior public histories affect the ex-ante value at the prior, and apply the framework to canonical applications.

The belief-space representation (\cref{prop:value-representation}) reduces the seller's ex-ante problem to an information design problem on the space of beliefs. Because the good is allocated only once (\cref{ass:single-shot}), each branch of a public experiment contributes one continuation value. Once we know what the seller can earn from each date and posterior, the remaining choice is how the prior is split across posteriors. In this sense, the dynamic mechanism problem has a static representation: the seller chooses a distribution of posterior beliefs, and each posterior is assigned the continuation value attainable from that point on.

\Cref{thm:nogain-delta-eps} shows that the non-posterior channel leaves the seller's ex-ante value unchanged precisely when a single shadow value function of the state both dominates every continuation value the richer history can attain and meets the posterior-only optimum at the prior. When the condition holds, \Cref{prop:collapse-structure} gives an implementable replacement for the optimal mechanism, a terminal one that discloses once to split the prior into finitely many posteriors and thereafter conditions on the belief alone rather than on the full public history, at the same ex-ante revenue.

The criterion is stated through continuation values, but checking it does not require solving the whole dynamic continuation problem (\cref{prop:criterion transform}). With a single-shot resource, the dynamic problem has the structure of a stopping problem, which reduces the comparison between richer and posterior-only histories to a within-date one.

The applications show why the answer depends on where the mechanism uses information. In monopoly screening, the public cost or demand shock changes payoffs but does not reveal the buyer's type. The relevant comparison reduces to known types: state-contingent and state-independent mechanisms have the same ex-ante value if and only if they coincide at known types, $(S_0)$-almost surely (\cref{prop:screen-collapse}). In certification, the seller first discloses quality to obtain approval and then faces a demand shock. The posterior at which trade occurs is chosen by the certification problem itself. The value of conditioning on demand is therefore evaluated at the approval posterior, not at known qualities. As a result, ex-ante pricing and state-contingent pricing coincide only in the knife-edge case identified in (\cref{prop:cert-collapse}).

\paragraph{Related work}
The question of when dynamic mechanisms reduce to history-independent objects is long-standing. \citet{Stokey1979} first asked, for monopoly intertemporal pricing, whether dynamic adjustment improves on a single commitment price; \citet{courty2000} provided the canonical opposite case in sequential screening, where dynamic contracts strictly dominate static ones. \citet{KrahmerStrausz} raise the comparison to a representation level: every sequential screening problem is equivalent, via a type-space augmentation, to a static screening problem.

Technically, the duality through which we translate equivalence into a single continuous affine support on the belief space is closest to \citet{DworczakKolotilin2024}, who show that the optimal dual variable in Bayesian persuasion is a supergradient of the concave closure of the objective at the prior. Concavification on the belief space originates with \citet{kamenica2011bayesian} and \citet{bergemann2016information}. The infinite-dimensional belief space is handled by standard analytic \citep{BertsekasShreve,Kechris} and concave-envelope techniques \citep{AliprantisBorder}.

A growing recent literature identifies environments in which the optimal dynamic mechanism reduces, in equilibrium or in ex-ante value, to a simpler object. \citet{EsoSzentes2017} establish a dynamic irrelevance theorem: after orthogonalisation, an agent's post-contract private information generates no rent. \citet{BoardPycia2014} show that, with an outside option, a durable-goods monopolist charges a constant monopoly price in every period. \citet{BergemannCastroWeintraub2020} give a primitive necessary-and-sufficient condition for the static contract to be optimal in sequential screening under ex-post participation. \citet{BanchioYang2021} show that, with limited commitment and persistent private types, the static price path is the unique equilibrium outcome. \citet{BuehlerEschenbaum2021} show that pricing dynamics do not emerge once trading-up opportunities from the static optimum are exhausted. \citet{AshlagiDaskalakisHaghpanah2023} characterise when the optimal mechanism in single-buyer sequential auctions collapse to ignore-history monopoly pricing. \citet{DovalSmolin2026} show that, in calibrated single-agent mechanism design, every implementable outcome is generated by a date-$0$ disclosure followed by a state-independent allocation rule. These results identify specific conditions under which the dynamic mechanism collapses; the characterisation here, via a continuous affine shadow value that supports the posterior-only benchmark  and dominates every continuation value, provides a unifying belief-space perspective on such reductions.

\paragraph{Roadmap} The remainder of the paper is organized as follows. Section~\ref{sec:prelim} introduces the model and the representation. Section~\ref{sec:collapse-stat} provides the main theorem and results. Section~\ref{sec:Empirical-interpretation} presents the applications.


\section{Model}\label{sec:prelim}
\subsection{Preliminaries}
\paragraph{Environment.} Time is discrete, with dates $t\in\T:=\{0,1,2,\dots\}$. There is a seller (designer) and a set of potential buyers for a resource. The payoff-relevant uncertainty is defined on a Borel probability space $(\Omega,\F,\mathbb P)$ together with the primitive state $\theta:\Omega\to\Theta$ where $\Theta$ is a compact Polish space. The set $\Delta(\Theta)$ denotes the set of probability measures on $\Theta$, equipped with the weak topology. At each date \(t\), public histories are represented by a public state $Z_t:\Omega\to E_t$ where \(E_t\) is a Polish space and $\F_t=\sigma(Z_t)$ is an increasing filtration of public-history $\sigma$-algebras with $\F_0$ trivial. A date-\(t\) continuation mechanism rule is an \(\F_t\)-measurable map $\Gamma_t:\Omega\to\mathcal H_t$ where \(\mathcal H_t\) is a Polish space of continuation mechanisms.\footnote{For example, $\Gamma_t$ may specify a posted price, a menu, an auction, or a timing lottery. A continuation mechanism $\Gamma_t$ is $\F_t$-measurable if its realization depends only on the public history $\F_t$.}

\paragraph{Signal and experiment.} We assume that the date-0 public experiment is freely designable within the class of finite public signals. Hence every finite Bayes-plausible distribution over posteriors can be induced by some admissible public experiment. Conversely, by the posterior martingale (see \cref{lem:posterior-martingale}), every public experiment induces a Bayes-plausible posterior distribution.

\begin{assumption}[Single-shot resource]\label{ass:single-shot}
The resource is economically single-shot: it can be allocated at most once. If the resource is assigned to some buyers and transfers are made, then no later date $t'>t$ can generate additional surplus.\footnote{If  Assumption \ref{ass:single-shot} does not hold, the problem may degenerate into a simple problem of finding the pointwise optimal continuation mechanism at each date $t$, the intertemporal problem loses the single-shot structure that motivates our analysis.}
\end{assumption}

\paragraph{Timeline.} Let $\tau\in\T\cup\{\varnothing\}$ denote the date at which the resource is actually allocated. If $\tau=\varnothing$, no sale ever occurs. At each date \(t\) and after any public state \(Z_t\), the seller chooses a continuation mechanism \(\Gamma_t\in\mathcal H_t\); past announcements about future continuation mechanisms are not binding. We record the resulting \emph{dynamic mechanism calendar} as $\mathcal M=\{(Z_1,\Gamma_1),(Z_2, \Gamma_2), ...\}$. Let \(R(\mathcal M)\) denote the supremum of the seller's ex-ante payoff over all sequentially rational Bayesian equilibria of the dynamic game associated with the calendar $\mathcal M$.

\begin{figure}[h]
\centering
\begin{tikzpicture}[
    >=Latex,
    font=\small,
    date/.style={
        draw,
        rounded corners,
        align=center,
        minimum width=2.5cm,
        minimum height=1.1cm
    },
    event/.style={
        draw,
        rounded corners,
        align=center,
        minimum width=2.8cm,
        minimum height=1.1cm,
        fill=gray!10
    },
    arrow/.style={->, thick}
]

\node[date] (d0) {
    Date \(0\)\\
    \(Z_0\), choose \(\Gamma_0\)
};

\node[date, right=1.4cm of d0] (d1) {
    Date \(1\)\\
    \(Z_1\), choose \(\Gamma_1\)
};

\node[right=1.1cm of d1] (dots) {
    \(\cdots\)
};

\node[date, right=1.1cm of dots] (dtau) {
    Date \(\tau\)\\
    \(Z_\tau\), choose \(\Gamma_\tau\)
};

\node[event, right=1.4cm of dtau] (alloc) {
    Resource allocated\\
    transfers made
};

\node[event, below=1.2cm of dtau] (nosale) {
    No sale ever occurs\\
    \(\tau=\varnothing\)
};

\draw[arrow] (d0.east) -- (d1.west);
\draw[arrow] (d1.east) -- (dots.west);
\draw[arrow] (dots.east) -- (dtau.west);
\draw[arrow] (dtau.east) -- (alloc.west);

\draw[arrow, dashed] (dots.south) -- (nosale.west);

\node[above=0.45cm of dots, align=center, font=\scriptsize] {
    Mechanism calendar recording: \(\mathcal M=\{M_t=(Z_t,\Gamma_t)\}_{t\in\mathbb T}\)
};


\end{tikzpicture}
\caption{The dynamic mechanism calendar evolving with a single-shot resource.}
\label{fig:model-timeline}
\end{figure}

\subsection{Seller's optimal revenue}\label{subsec:reduced-form}

To characterize \(R(\mathcal M)\), we identify the seller's payoff at each date of the game. That payoff is the value the seller can attain in the continuation problem itself, determined by its equilibria rather than defined recursively from later dates. The definitions that follow only make this object precise.

\paragraph{Continuation equilibria.} Fix a date \(t\), a public state \(z\in E_t\), and a continuation mechanism \(h\in\mathcal H_t\). Let $\mathcal B_t(z,h)$ denote the set of strategy--belief profiles of the continuation game that starts at date \(t\) after public state \(z\), when the chosen continuation mechanism is \(h\). Let $\mathcal E_t(z,h)\subseteq \mathcal B_t(z,h)$ denote the subset of continuation profiles that are Bayesian equilibria of this date-\(t\) continuation game. For each \(b\in\mathcal E_t(z,h)\), let
\[
\pi_t(\omega,z,h,b)\in\mathbb R
\]
be the seller's state-contingent payoff function under \(b\). For a continuation mechanism rule \(\Gamma_t\), the induced random payoff correspondence is
\[
\mathcal R_t^{\Gamma_t}(\omega)
:=
\bigl\{
\pi_t(\omega,Z_t(\omega),\Gamma_t(\omega),b),b \in \mathcal E_t(Z_t(\omega),\Gamma_t(\omega))
\bigr\}.
\]

\begin{assumption}[Analytic equilibrium]
\label{ass:analytic-equilibrium}
For every date $t\in\mathbb T$, every continuation mechanism $h\in\mathcal H_t$, and every public state $z\in E_t$:
\begin{enumerate}[label=\textup{(\roman*)}]
    \item\label{item:nonempty}
    The continuation equilibrium profile set $\mathcal E_t(z,h)$ is nonempty.
    \item\label{item:analytic-graph}
    The graph of the equilibrium correspondence $\operatorname{Gr}(\mathcal E_t)
    :=
    \bigl\{(z,h,b)\in E_t\times\mathcal H_t\times\mathcal B_t :
    b\in\mathcal E_t(z,h)\bigr\},$
    is an analytic subset of $E_t\times\mathcal H_t\times\mathcal B_t$.
    \item\label{item:meas-payoff}
    The payoff function
    $\pi_t$ is jointly Borel measurable.
    \item\label{item:bounded}
    The payoff correspondence $\mathcal R_t^{\Gamma_t}$ is 
    uniformly bounded above.\footnote{Under \ref{item:nonempty}--\ref{item:meas-payoff}, the payoff correspondence $\mathcal R_t^{\Gamma_t}$ is automatically nonempty-valued and has analytic graph, so these properties need not be assumed separately. Condition~\ref{item:bounded} is the only independent content in \textup{(iv)}: it is an economic requirement that equilibrium revenues are finite.}
\end{enumerate}
\end{assumption}

\begin{remark}
    Condition~\ref{item:analytic-graph} is satisfied in two leading cases that cover most applications. First, if $\mathcal B_t(z,h)$ is a compact metric space and the best-reply correspondence is upper hemicontinuous with nonempty compact values, then $\operatorname{Gr}(\mathcal E_t)$ is closed, hence Borel, hence analytic. Second, if the strategy space is finite (e.g.\ finite auctions or
    finite menus), $\mathcal E_t(z,h)$ is a finite union of points for each $(z,h)$, so its graph is a countable union of Borel sets and thus analytic. In both cases \ref{item:analytic-graph} follows without
    additional argument.
\end{remark}

\paragraph{Continuation value.} For a continuation mechanism $\Gamma$, let 
$\Sel(\mathcal R_t^{\Gamma_t})$ denote the set of universally measurable payoff selections of $\mathcal R_t^{\Gamma_t}$ (see \cref{lem:selection-exists}). Public history $Z_t$ induces a posterior $S_t:\Omega\to\Delta(\Theta)$ which summarizes all public learning about the primitive state $\theta$ (see \cref{lem:posterior-martingale}). Let $\mathbb P_t^\mu(\cdot)$ be a version of $\mathbb P(\cdot\mid S_t(\omega)=\mu)$, which denotes the conditional law of the state $\omega$ given the public posterior $\mu$. \emph{Conditioning on} the public continuation state \(Z_t\), the date-\(t\) \emph{continuation value} is\footnote{the definition is expectation-based: the object of interest is the seller's date-$t$ conditional expected equilibrium payoff at a given posterior, not a pathwise conditional essential supremum over revenues. Whenever a continuation value $U_t^{\mathcal G}$ admits multiple measurable versions, we fix once and for all a bounded upper semianalytic version and continue to denote it by the same symbol.} 
\[
U_t^{\sigma(Z_t)}(\mu)
:=
\sup\Bigg\{
\int_\Omega \rho(\omega)\,\mathbb P_t^\mu(d\omega):
\Gamma \text{ is } \sigma(Z_t)\text{-measurable},
\ 
\rho\in \Sel\big(\mathcal R_t^{\Gamma_t}\big)
\Bigg\},
\qquad
\mu\in\Delta(\Theta).
\]
\begin{remark}[Continuation value]\label{Continuation value}
The continuation value \(U_t^{\sigma(Z_t)}\) is the seller's maximal conditional expected equilibrium payoff in the date-\(t\) continuation mechanism problem, given the public state \({Z_t}\) on which the continuation rule may condition. It is not a Bellman continuation value, but rather the equilibrium payoff from the continuation problem starting at date \(t\).
\end{remark}

\paragraph{Equilibrium-branch representation.} Let $\pi=\sum_{k=1}^K \alpha_k\delta_{\mu_k}$ be a finite-support posterior distribution induced by a public experiment,\footnote{Here \(\pi\) denotes a probability distribution over posterior beliefs; this notation is distinct from the payoff function \(\pi_t\).} where \(\alpha_k\) is the probability that posterior \(\mu_k\) is realized. For each \(k\), choose \(t_k\in\mathbb T\) and an \(\varepsilon\)-optimal selector \((\Gamma_k,\rho_k)\) for
\(U_{t_k}^{\mathcal F_{t_k}}(\mu_k)\), that is,
\[
\int_\Omega \rho_k(\omega)\,
\mathbb P_{t_k}^{\mu_k}(d\omega)
\ge
U_{t_k}^{\mathcal F_{t_k}}(\mu_k)-\varepsilon .
\]

\noindent\emph{Lower bound.} Under \cref{ass:analytic-equilibrium}, every payoff selection admits a universally measurable equilibrium lifting (see \cref{lem:measurable-lifting}). So the sequential pasting argument (see \cref{lem:sequential-pasting}) applied, these continuation selectors can be assigned after the posterior realizations of the date-0 public experiment to form a sequentially rational Bayesian equilibrium of the resulting mechanism calendar. The seller's ex-ante payoff in this equilibrium is at least
\[
R(\mathcal M) \geq \sum_{k=1}^K
\alpha_k U_{t_k}^{\mathcal F_{t_k}}(\mu_k)
-\varepsilon .
\]
This inequality states that the dynamic mechanism calendar can implement, up to \(\varepsilon\), the probability-weighted average of the posterior-contingent date-\(t_k\) continuation values generated after the public experiment.\footnote{The split used in the representation is not a recursive Bellman split. It is a split over complete continuation-equilibrium branches. For each posterior realization, the selected object is a full sequentially rational continuation profile of the game starting from the corresponding public continuation state.}

\bigskip
\noindent\emph{Upper bound.} For any function \(q:\Delta(\Theta)\to\mathbb R\), define its finite concavification by
\[
(\operatorname{conc}_f q)(\mu)
:=
\sup_{\pi\in\mathcal P_f(\Delta(\Theta)):
\int_{\Delta(\Theta)}\nu\,\pi(d\nu)=\mu}
\int_{\Delta(\Theta)}q(\nu)\,\pi(d\nu).
\]

\begin{assumption}[Regularity]\label{ass:regularity}
The finite concavification
$\operatorname{conc}_f\!\big[\sup_{t\in\mathbb T}U_t^{\mathcal G_t}\big]$
is upper semicontinuous at the prior $S_0$.
\end{assumption} 
\noindent Under \cref{ass:regularity},\footnote{\cref{ass:regularity} is the exact property used in the proofs of \cref{lem:upper-bound,lem:concf-conc}: the concavified envelope must not jump upward at $S_0$. It is implied by standard game-theoretic primitives. See \cref{app:regularity-sufficient} for a layered statement of primitive sufficient conditions.} the seller's ex-ante equilibrium payoff is bounded above by the finite concave envelope of the date-wise upper envelope of full-history continuation values (see \cref{lem:upper-bound}); the envelope is evaluated at the prior belief \(\mu_0\), where \(S_0(\omega)\equiv \mu_0\) for trivial \(\mathcal F_0\). 

\medskip
Combining this upper bound with the lower bound gives the belief-space representation of the seller's optimal ex-ante payoff:

\begin{proposition}[Equilibrium-branch representation]
\label{prop:value-representation}
Suppose \cref{ass:single-shot,ass:analytic-equilibrium,ass:regularity} holds, the seller's ex-ante payoff satisfies
\[
\sup_{\mathcal M} R(\mathcal M)
= 
\Big(\operatorname{conc}_f
\big[\sup_{t\in\mathbb T} U_t^{\sigma(Z_t)}\big]\Big)(S_0).
\]
\end{proposition}

Two features make this representation available where it usually is not. The resource is allocated at most once, so the seller's payoff is realized at a single date and no surplus is double-counted across dates (\cref{ass:single-shot}). The posterior is a martingale, so beliefs are linked across dates only through Bayes plausibility, which is the only restriction on the posterior distributions a date-0 experiment can induce. The date-wise inputs \(U_t^{\sigma(Z_t)}\) are continuation equilibrium values rather than Bellman continuation values, so the branchwise selectors are pasted into a single sequentially rational equilibrium of the induced calendar (\cref{lem:measurable-lifting,lem:sequential-pasting}), with no separate consistency condition to verify.


\subsection{The value beyond posterior }\label{subsec:representation.}
\paragraph{State decomposition.} As mentioned above, the posterior $S_t$ is sufficient for inference about $\theta$. However, two realized public histories may induce the same posterior $S_t$ and still lead to different date-$t$ continuation problems.\footnote{Examples include eligibility states, promised utilities, inventories, reputations, or other public variables that do not further change beliefs about \(\theta\), but may affect the continuation problem.} To capture this distinction, we impose the following decomposition of the public state. Without loss of generality, we rewrite the date-$t$ public state as
\[
Z_t:=(S_t,Y_t),
\]
where $S_t$ is the posterior and $Y_t$ denotes the non-posterior public continuation state.\footnote{Non-posterior public state $Y_t$ does not change beliefs about $\theta$ beyond what is already encoded in $S_t$, but it may remain payoff-relevant for the induced date-$t$ continuation mechanism problem. We provided a specific definition (\cref{eq:Non-posterior}) in \cref{sec:screening-application}.} Posterior $S_t$ summarizes public learning about the primitive state, whereas $(S_t,Y_t)$ is the full public state relevant for the date-$t$ mechanism problem. The state decomposition allows us to compare different public states at date $t$. Let $\mathcal G$ be a public state satisfying
\[
\sigma(S_t)\subseteq \mathcal G\subseteq \sigma(S_t,Y_t).
\]
That is, $\mathcal G$ summarizes the public states on which the seller is allowed to condition the date-$t$ continuation mechanism. 

\paragraph{Non-posterior value.} Fix a date \(t\), a posterior \(\mu\in\Delta(\Theta)\), and a public state \(\mathcal G\), a \(\mathcal G\)-admissible selector at \(\mu\) is a pair \((\Gamma_t,\rho)\) such that \(\Gamma_t\) is \(\mathcal G\)-measurable and $\rho\in \Sel(\mathcal R_t^{\Gamma_t})$. A selector \((\Gamma_t,\rho)\) is \(\mathcal G\)-\emph{optimal} at \(\mu\) if
\[
\int_\Omega \rho(\omega)\,\mathbb P_t^\mu(d\omega)
=
U_t^{\mathcal G}(\mu).
\]

Allowing richer public states $\{\G_t\}$ and a larger class of continuation mechanisms expands each date-$t$ continuation value $U_t$. When $\mathcal G=\sigma(S_t)$, the seller may condition only on the posterior belief; when $\mathcal G=\sigma(Z_t)$, the seller may condition on the realized public full-history, including the non-posterior continuation state $Y_t$. The gap between $U_t^{\sigma(S_t)}$ and $U_t^{\sigma(Z_t)}$ measures the incremental date-$t$ continuation value of non-posterior public states beyond beliefs. 

\medskip
Given public state $\sigma(S_t)\subseteq \mathcal G\subseteq \sigma(S_t,Y_t)$ and $Y_t$, we say that the selector collapse holds from \(\sigma(S_t,Y_t)\) to \(\mathcal G\) if, for every \(\mu\), there exists a \(\sigma(S_t,Y_t)\)-optimal selector \((\Gamma_t,\rho)\) at \(\mu\) such that \(\Gamma_t\) is \(\mathcal G\)-measurable; and the continuation value collapse from \(\sigma(S_t,Y_t)\) to \(\mathcal G\) if
\[
U_t^{\sigma(S_t,Y_t)}(\mu)=U_t^{\mathcal G}(\mu)
\qquad
\forall\,\mu\in\Delta(\Theta).
\]
Continuation value collapse is weaker than selector collapse; the two coincide under additional structure.\footnote{In \cref{subsec:condi-selec-coll}, we record four sufficient conditions under which selector collapse to \(\mathcal G\) can be verified. } Date-wise selector collapses at every date is a sufficient  condition for the irrelevance of a richer public state. Its converse does not generally hold. We aim to characterize the necessary and sufficient conditions for the latter.

\section{Results}\label{sec:collapse-stat}
\subsection{State redundancy}

When does a richer public state strictly enhance the seller’s optimal revenue? The seller gains no additional ex-ante value if and only if there exists an affine \emph{shadow value function} $\lambda$ on the belief space that simultaneously supports the posterior benchmark at $S_0$ and majorizes all continuation values $U_t^{\G_t}$.

\begin{theorem}[No gain]
\label{thm:nogain-delta-eps}
The richer public state $\mathcal{G}_t$ with $\sigma(S_t) \subseteq \mathcal{G}_t \subseteq \sigma(S_t, Y_t)$, generates no additional value at the prior, if and only if, for every $\varepsilon > 0$, there exists a continuous shadow value function $\lambda_\varepsilon \colon \Theta \to \mathbb{R}$ such that:
\begin{enumerate}[label=\textup{(\roman*)}]
    \item \label{cond:posterior-support}\textup{(Posterior support)}\enspace
    \[
    \sup_{t \in \mathbb{T}} U_t^{\sigma(S_t)}(\mu) \le \int_\Theta \lambda_\varepsilon(\theta) \, \mu(d\theta) \quad \textup{for all } \mu \in \Delta(\Theta);
    \]
    \item \label{cond:tightness}\textup{(Tightness at the prior)}\enspace
    \[
    \int_\Theta \lambda_\varepsilon(\theta) \, S_0(d\theta) \le \left(\operatorname{conc}_f \left[\sup_{t \in \mathbb{T}} U_t^{\sigma(S_t)}\right]\right)(S_0) + \varepsilon.
    \]
    \item \label{cond:richer-cap}\textup{(Richer-state cap)}\enspace
    \[
    U_t^{\mathcal{G}_t}(\mu) \le \int_\Theta \lambda_\varepsilon(\theta) \, \mu(d\theta) \quad \textup{for all } \mu \in \Delta(\Theta) \textup{ and all } t \in \mathbb{T};
    \]
\end{enumerate}
\end{theorem}

\noindent\emph{Idea of proof.}\enspace
The argument is a duality. By the value representation (\cref{prop:value-representation}), the prior value under either state is the finite concavification of the corresponding envelope, so no gain means the richer and posterior-only envelopes share the same concavified value at the prior; as a richer state only raises the envelope, one direction is automatic and only the reverse must be shown. By duality, it holds exactly when an affine shadow value $\lambda$ supports the posterior benchmark $g$ at $S_0$ and majorizes the richer envelope, which are \cref{cond:posterior-support,cond:tightness,cond:richer-cap}. Such a $\lambda$ exists because the concavified value is concave and, by \cref{ass:regularity}, upper semicontinuous at $S_0$, so a supporting hyperplane is available there. Given $\lambda$, the richer value at the prior cannot exceed that of $g$, since $\lambda$ lies above the richer envelope and beliefs average back to $S_0$; conversely, that supporting hyperplane is the required $\lambda$.

\medskip
\Cref{thm:nogain-delta-eps} characterizes when the non-posterior state leaves the seller's ex-ante value unchanged. A richer history can raise the continuation value above the posterior-only benchmark, but only at beliefs the optimal disclosure from $S_0$ never induces. At the beliefs disclosure does reach, the richer history adds nothing beyond the benchmark, so the extra value is never collected and the ex-ante value is unaffected. Non-posterior histories therefore shape the continuation mechanism without moving ex-ante revenue.

\medskip
\begin{figure}[H]
  \centering

  \def\legendline#1{%
    \tikz[baseline=-0.6ex]\draw[#1] (0,0)--(0.65cm,0);%
  }

    \makebox[\textwidth][c]{%
    \fbox{%
      \begin{minipage}{0.5\textwidth}
        \centering
        \small
        \legendline{thick, blue}\,$g(\mu)$\quad
        \legendline{thick, red, dashed}\,$\widehat g(\mu)$\quad
        \legendline{thick, gray}\,$\operatorname{conc} g(\mu)$\quad
        \legendline{thick, dashdotted, green!50!black}\,$\ell(\mu)$%
      \end{minipage}%
    }%
  }

  \vspace{0.6em}
  
  \begin{tikzpicture}
    \begin{axis}[
      width=0.7\textwidth,
      height=4.2cm,
      xlabel={Belief $\mu$},
      ylabel={Revenue},
      xmin=0, xmax=1,
      ymin=0.18, ymax=0.55,
      axis lines=left,
      ticklabel style={font=\small},
      label style={font=\small},
      grid=none,
      xtick=\empty,
      ytick=\empty
    ]

      \addplot[thick, blue, domain=0:1, samples=200]
        {0.6*x*(1 - x) + 0.197};

      \addplot[thick, gray, domain=0:1, samples=200]
        {0.6*x*(1 - x) + 0.2};

      \addplot[thick, red, dashed, domain=0:1, samples=200]
        {0.6*x*(1 - x) + 0.203 + 0.08*max(x-0.7,0)/0.3};

      \addplot[thick, dashdotted, domain=0:1, samples=200, green!50!black]
        {0.24*x + 0.257};

      \def\Szero{0.3}
      \def\gSzero{0.326} 

      \addplot[dotted, gray]
        coordinates {(\Szero,0.18) (\Szero,0.55)};

      \addplot[only marks, mark=*, mark options={fill=gray}, gray]
        coordinates {(\Szero,\gSzero)};

      \node[anchor=south west, font=\small]
        at (axis cs:\Szero+0.02,\gSzero+0.04) {$S_0$};

        \node[anchor=north west, font=\scriptsize, text=black]
        at (axis cs:\Szero+0.02,0.285) {$g$ concave};

    \end{axis}
  \end{tikzpicture}
  \begin{tikzpicture}
    \begin{axis}[
      width=0.72\textwidth,
      height=4.6cm,
      xlabel={Belief $\mu$},
      ylabel={Revenue},
      xmin=0, xmax=1,
      ymin=0.34, ymax=0.54,
      axis lines=left,
      ticklabel style={font=\small},
      label style={font=\small},
      grid=none,
      xtick=\empty,
      ytick=\empty
    ]

      \addplot[
        thick,
        blue
      ] coordinates {
        (0.00,0.40)
        (0.10,0.37)
        (0.20,0.50)
        (0.35,0.41)
        (0.50,0.38)
        (0.65,0.42)
        (0.80,0.50)
        (0.90,0.39)
        (1.00,0.40)
      };

      \addplot[
        thick,
        red,
        dashed
      ] coordinates {
        (0.00,0.42)
        (0.10,0.39)
        (0.20,0.50)
        (0.35,0.43)
        (0.50,0.42)
        (0.65,0.45)
        (0.80,0.50)
        (0.90,0.43)
        (1.00,0.42)
      };

      \addplot[
        thick,
        gray
      ] coordinates {
        (0.00,0.40)
        (0.20,0.50)
        (0.80,0.50)
        (1.00,0.40)
      };

      \addplot[
        thick,
        dashdotted,
        green!50!black,
        domain=0:1,
        samples=2
      ] {0.50};

      \def\Szero{0.50}

      \addplot[dotted, gray]
        coordinates {(\Szero,0.34) (\Szero,0.54)};

      \addplot[
        only marks,
        mark=*,
        mark options={fill=gray},
        gray
      ] coordinates {(\Szero,0.50)};


      \node[anchor=south west, font=\small]
        at (axis cs:\Szero+0.02,0.462) {$S_0$};

      \node[anchor=north west, font=\scriptsize, text=black]
        at (axis cs:\Szero+0.02,0.385) {$g$ non-concave};


    \end{axis}
  \end{tikzpicture}
  
  \caption{The posterior-only envelope $g(\mu):=\sup_t U_t^{\sigma(S_t)}(\mu)$
(solid blue), its concavification $\operatorname{conc} g(\mu)$
(solid gray), the richer state envelope
$\widehat g(\mu):=\sup_t U_t^{\mathcal G_t}(\mu)$
(dashed), and an affine shadow value
$\ell(\mu):=\int_\Theta \lambda_\varepsilon(\theta)\,\mu(d\theta)$
(dash--dotted) that supports $\operatorname{conc} g$ at $S_0$ and dominates
$\widehat g$ on $\Delta(\Theta)$.}
\label{fig:shadow-value}
\end{figure}

\subsection{Mechanism replacement} 
The posterior-only benchmark induces a reduced calendar: a finite date-0 experiment over posteriors, followed on each branch \(\mu\) by a date \(t(\mu)\) and a \(\sigma(S_{t(\mu)})\)-admissible continuation selector. This motivates the following definition.

\begin{definition}[Terminal]
\label{def:collapse}

A mechanism calendar \(\widetilde{\mathcal M}\) is called \emph{terminal} if a date-0 public experiment induces a posterior \(\mu\in\Delta(\Theta)\), and the calendar then specifies a date \(t(\mu)\in\mathbb T\) together with a selector $(\Gamma^\mu,\rho^\mu)$ that is \(\sigma(S_{t(\mu)})\)-admissible at \(\mu\).

A mechanism calendar \(\mathcal M\) \emph{collapses to a terminal mechanism} at \(S_0\) if, for every \(\varepsilon>0\), there exists a terminal mechanism calendar \(\widetilde{\mathcal M}_\varepsilon\) such that
\[
R(\widetilde{\mathcal M}_\varepsilon)
\ge
R(\mathcal M)-\varepsilon .
\]
\end{definition}


\begin{proposition}[Collapse criterion]\label{prop:collapse-structure}
The full-history mechanism calendar collapses to a terminal mechanism at $S_0$ if and only if, for every $\varepsilon>0$, there exists a continuous shadow value function $\lambda_\varepsilon\colon\Theta\to\mathbb R$ satisfying \ref{cond:tightness}, and
\begin{equation}
U_t^{\mathcal F_t}(\mu)\le\int_\Theta\lambda_\varepsilon(\theta)\,\mu(d\theta)
\qquad\forall\,\mu\in\Delta(\Theta),\ \forall\,t\in\mathbb T.
\tag{Terminal Support}\label{eq:terminal-support}
\end{equation}
\end{proposition}

\noindent\emph{Idea of proof.}\enspace
A terminal calendar is a date-$0$ public experiment followed on each branch by a $\sigma(S_t)$-admissible continuation selector, so its value is at most the concavification of the posterior benchmark $g$; this bound is attained by pasting near-optimal branchwise selectors into an equilibrium calendar (\cref{lem:measurable-lifting,lem:sequential-pasting}). The full-history calendar, by contrast, earns the concavification of the richer envelope (\cref{prop:value-representation}). Collapse is therefore the equality of the two at the prior, which is exactly no gain for the full public state $\mathcal F_t$, so the claim is \cref{thm:nogain-delta-eps} applied with $\mathcal G_t=\mathcal F_t$, the richer-state cap becoming \cref{eq:terminal-support}; the richer-state cap \ref{cond:richer-cap} implies the posterior support \ref{cond:posterior-support}.

\medskip
Where \cref{thm:nogain-delta-eps} concerns the seller's revenue, \cref{prop:collapse-structure} concerns the rule that earns it. In a full-history calendar the continuation rule may track the entire public record, so two histories that leave the same belief about the state can still be met with different rules when they differ in the non-posterior public state. The proposition shows that this dependence can be dispensed with exactly when the \cref{eq:terminal-support} condition holds, that is, when a single shadow value lies above every continuation value the full history can generate while still meeting the optimal value at the prior. In that case the full-history calendar collapses to a terminal mechanism, in which a single public experiment at the outset splits the prior into finitely many posterior beliefs and every continuation rule thereafter depends on the belief alone, and the seller collects the same ex-ante revenue. This replacement is ex-ante; it does not assert that every optimal full-history continuation rule is posterior-measurable after every history. Non-posterior histories still reshape the continuation rule, but only at beliefs the optimal disclosure never reaches, and so never where revenue is made.

\paragraph{A diagnostic.}\label{para:diagnostic} Conditions \ref{cond:posterior-support} and \ref{cond:tightness} pin down a \emph{shadow value}: a state valuation $\lambda\colon\Theta\to\mathbb R$ that prices no belief below the posterior-only benchmark $g:=\sup_t U_t^{\sigma(S_t)}(\mu)$ and, at the prior, costs at most the optimum (up to $\varepsilon$). The non-posterior gain region collects the belief-value pairs lying above every such price,
\[
\mathcal N(S_0)
:=
\Big\{(\mu,r)\in\Delta(\Theta)\times\mathbb R:\
r>\sup_{\lambda}\int_\Theta\lambda(\theta)\,\mu(d\theta)\Big\},
\]
the supremum over continuous $\lambda$ meeting condition \ref{cond:posterior-support} and \ref{cond:tightness}. By construction it depends only on $g$ and $S_0$, not on the richer histories. Collapse fails as soon as $\widehat g:=\sup_t U_t^{\F_t}(\mu)$ enters $\mathcal N(S_0)$ at some belief

\begin{corollary}[Diagnostic]\label{prop:diagnostic}
If the full-history calendar collapses to a terminal mechanism at $S_0$, then
for every $\mu\in\Delta(\Theta)$,
\[
(\mu,\widehat g(\mu))\notin\mathcal N(S_0).
\]
\end{corollary}

\medskip
\begin{figure}[H]
  \centering

  \begin{minipage}[t]{0.48\textwidth}
    \vspace{0pt}
    \centering
    \begin{tikzpicture}
      \begin{axis}[
        width=\textwidth,
        height=4.5cm,
        axis lines=left,
        xmin=0, xmax=1,
        ymin=0.35, ymax=0.7,
        xtick=\empty,
        ytick=\empty,
        xlabel={Belief $\mu$},
        ylabel={Revenue},
        label style={font=\small},
        ticklabel style={font=\small},
        legend style={font=\scriptsize, at={(0.5,1.03)}, anchor=south, legend columns=4},
        legend cell align=left,
        grid=none,
        clip=false
      ]

        \addplot[thick, blue, domain=0:1, samples=200]
          {min(0.2*x + 0.4, -0.2*x + 0.6)};
        \addlegendentry{$g(\mu)$}

        \addplot[thick, red, dashed, domain=0:1, samples=200]
          {min(0.2*x + 0.4, -0.2*x + 0.6035) + 0.08*exp(-80*(x-0.2)^2)};
        \addlegendentry{$\widehat g(\mu)$}

        \addplot[thick, dashdotted, gray, domain=0:1, samples=200]
          {0.2*x + 0.4035};
        \addlegendentry{$\ell$}

        \addplot[thick, dotted, green!60!black, domain=0:1, samples=200]
          {-0.2*x + 0.605};
        \addlegendentry{$\ell'$}

        \def\Szero{0.5}
        \def\gSzero{0.5}

        \addplot[dotted, gray, forget plot]
          coordinates {(\Szero,0.35) (\Szero,0.7)};

        \addplot[only marks, mark=*, mark size=2pt, blue, forget plot]
          coordinates {(\Szero,\gSzero)};

        \node[anchor=south west, font=\small]
          at (axis cs:\Szero+0.02,\gSzero+0.01) {$S_0$};

        \node[anchor=north west, font=\small]
          at (axis cs:0.02,0.69) {(a)};

      \end{axis}
    \end{tikzpicture}
  \end{minipage}
  \hfill
  \begin{minipage}[t]{0.48\textwidth}
    \vspace{0pt}
    \centering
    \begin{tikzpicture}
      \begin{axis}[
        width=\textwidth,
        height=4.5cm,
        axis lines=left,
        xmin=0, xmax=1,
        ymin=0.35, ymax=0.7,
        xtick=\empty,
        ytick=\empty,
        xlabel={Belief $\mu$},
        ylabel={Revenue},
        label style={font=\small},
        ticklabel style={font=\small},
        legend style={font=\scriptsize, at={(0.5,1.03)}, anchor=south, legend columns=3},
        legend cell align=left,
        grid=none,
        clip=false
      ]

        \addplot[thick, blue, domain=0:1, samples=200]
          {min(0.2*x + 0.4, -0.2*x + 0.6)};
        \addlegendentry{$g(\mu)$}

        \addplot[thick, red, dashed, domain=0:1, samples=200]
          {min(0.2*x + 0.4, -0.2*x + 0.6035) + 0.15*exp(-80*(x-0.2)^2)};
        \addlegendentry{$\widehat g(\mu)$}

        \addplot[thick, dashdotted, gray, domain=0:1, samples=200, forget plot]
          {0.2*x + 0.4035};

        \addplot[thick, dashdotted, gray, domain=0:1, samples=200, forget plot]
          {-0.2*x + 0.6035};

        \addplot[thick, dashdotted, gray, domain=0:1, samples=200, name path=env, forget plot]
          {max(0.2*x + 0.4035, -0.2*x + 0.6035)};

        \addplot[draw=none, domain=0:1, samples=2, name path=top, forget plot]
          {0.7};

        \addplot[fill=red!10, draw=none, forget plot]
          fill between[of=top and env];

        \addlegendimage{area legend, fill=red!10, draw=red!40}
        \addlegendentry{$\mathcal N(S_0)$}

        \def\Szero{0.5}
        \def\gSzero{0.5}

        \addplot[dotted, gray, forget plot]
          coordinates {(\Szero,0.35) (\Szero,0.7)};

        \addplot[only marks, mark=*, mark size=2pt, blue, forget plot]
          coordinates {(\Szero,\gSzero)};

        \node[anchor=north west, font=\small]
          at (axis cs:0.02,0.69) {(b)};

        \node[anchor=south east, font=\footnotesize, text=red!70!black]
          at (axis cs:0.98,0.59)
          {collapse fails if $\widehat g$ enters this region};

      \end{axis}
    \end{tikzpicture}
  \end{minipage}

  \caption{%
  Panel~(a) illustrates the Uniform Support condition: although \(\widehat g\) may exceed one support \(\ell:=\int_\Theta \lambda_\varepsilon(\theta) \, \mu(d\theta)\), collapse can still hold if another \(\ell':=\int_\Theta \lambda'_\varepsilon(\theta) \, \mu(d\theta)\) dominates it pointwise. Panel~(b) shows the gain region \(\mathcal N(S_0)\); entry of \(\widehat g\) into this region rules out uniform domination and hence implies failure of calendar collapse at \(S_0\).
  }
  \label{fig:gain-region}
\end{figure}

\subsection{Criterion transform}
\label{sec:Reducing the criteria}
The criteria of \cref{prop:collapse-structure,prop:diagnostic}
are stated through the continuation value envelopes, and are operational insofar as these envelopes can be evaluated. The next result shows that these criteria can be verified without solving the full continuation problem. Every condition in \cref{thm:nogain-delta-eps} can be checked on the static, within-date \emph{truncation value}, leaving the comparison between the rich and the posterior-only public state a within-period problem.

\paragraph{Truncation value.}
Fix $t\in\mathbb T$, a public state $z$, and a continuation mechanism $h\in\mathcal H_t$. The \emph{truncation} of the date-$t$ subgame $(z,h)$ replaces every public continuation history that reaches the start of date $t+1$ with the absorbing no-sale node, which yields the seller payoff $0$ and assigns each buyer the reservation payoff $0$; let $\mathcal E_t^{\circ}(z,h)$ denote its set of sequentially rational strategy-belief profiles. For an $\sigma(Z_t)$-measurable continuation rule $\Gamma_t$, the \emph{truncation payoff correspondence} is
\[
\mathcal V_t^{\Gamma_t}(\omega)
:=
\bigl\{
\pi_t\bigl(\omega,Z_t(\omega),\Gamma_t(\omega),b\bigr):
b\in\mathcal E_t^{\circ}\bigl(Z_t(\omega),\Gamma_t(\omega)\bigr)
\bigr\},
\]
and the \emph{truncation value} conditioning on a public state $\mathcal G_t$ with $\sigma(S_t)\subseteq\mathcal G_t\subseteq\sigma(Z_t)$ is
\[
V_t^{\mathcal G_t}(\mu)
:=
\sup\Bigl\{
\int_\Omega \rho(\omega)\,\mathbb P_t^\mu(d\omega):
\Gamma_t\text{ is }\mathcal G_t\text{-measurable},\
\rho\in\Sel\bigl(\mathcal V_t^{\Gamma_t}\bigr)
\Bigr\},
\qquad
\mu\in\Delta(\Theta).
\]
By construction $V_t^{\mathcal G_t}\ge 0$, it is bounded (\cref{ass:analytic-equilibrium}); it is a function of the posterior $\mu$, with $\mathcal G_t$ entering only as the conditioning richness of the continuation rule, and it involves no integration over the posterior process. Under the single-shot resource \cref{ass:single-shot}, the continuation value is the Snell envelope of the truncation values along the posterior process (see \cref{lem:stopping-form})
\begin{equation}\label{eq:stopping-form}
U_t^{\mathcal G_t}(\mu)=\max\Bigl\{V_t^{\mathcal G_t}(\mu),\
\int_\Omega U_{t+1}^{\mathcal G_{t+1}}\bigl(S_{t+1}(\omega)\bigr)\,\mathbb P_t^\mu(d\omega)\Bigr\}.
\end{equation}

\medskip
The option to defer lifts the static value $V_t$ to the Snell envelope $U_t$, raising it at beliefs from which waiting is profitable. An affine price, however, is neutral to this option. The shadow value $\lambda$ induces a price $\ell(\mu)=\int_\Theta\lambda(\theta)\,\mu(d\theta)$ that is affine in the belief; since beliefs evolve as a martingale, $\ell(S_t)$ is itself a martingale, and by optional sampling its value at the prior does not depend on the date at which the single allocation occurs. An affine price therefore dominates the Snell envelope $U_t$ exactly when it dominates the truncation values $V_t$, so the dual certificate of collapse is determined by $V_t$ alone (see \cref{lem:cap-equiv}).

\begin{proposition}[Criterion transform]\label{prop:criterion transform}
Fix $\varepsilon>0$. There is a continuous shadow value $\lambda_\varepsilon\colon\Theta\to\mathbb R$ satisfying \cref{cond:posterior-support,cond:tightness,cond:richer-cap} of \cref{thm:nogain-delta-eps} if and only if there is one satisfying the same three conditions with each continuation value $U_t^{\bullet}$ replaced by the corresponding truncation value $V_t^{\bullet}$. The collapse criterion and diagnostic are therefore verified on the static truncation value.
\end{proposition}
\begin{proof}
Write $\ell_\varepsilon(\mu)=\int_\Theta\lambda_\varepsilon(\theta)\,\mu(d\theta)$ for the affine price. Conditions \cref{cond:posterior-support} and \cref{cond:richer-cap} are the caps $\sup_t U_t^{\sigma(S_t)}\le\ell_\varepsilon$ and $\sup_t U_t^{\mathcal G_t}\le\ell_\varepsilon$ on $\Delta(\Theta)$; by \cref{lem:cap-equiv} an affine price caps a continuation value if and only if it caps the corresponding truncation value, so each is equivalent to its $V$-version. In \cref{cond:tightness}, the concavification identity of \cref{lem:cap-equiv} equates the benchmark $\bigl(\operatorname{conc}_f[\sup_t U_t^{\sigma(S_t)}]\bigr)(S_0)$ with $\bigl(\operatorname{conc}_f[\sup_t V_t^{\sigma(S_t)}]\bigr)(S_0)$, leaving the condition intact. Combining the three equivalences proves the claim.
\end{proof}

\paragraph{Connection to optimal stopping.} The reduction is the belief-space form of a standard fact. With a single-shot resource the allocation is a stopping decision, and the continuation value is the least superharmonic majorant of the truncation values \citep{peskir2006optimal}, whereas an affine function of the Bayes-plausible posterior is harmonic for the belief martingale \citep{aumann1995repeated,kamenica2011bayesian}. It is the same principle by which a martingale that dominates an American claim's exercise values dominates the claim itself, with the affine shadow value of \cref{thm:nogain-delta-eps} playing the role of that martingale on the belief space.

\section{Applications}\label{sec:Empirical-interpretation}

\subsection{Screening with non-posterior public state}
\label{sec:screening-application}
We ask what a monopolist gains by letting a screening mechanism depend on a payoff-relevant public state, such as a cost or demand shock, that affects profit but carries no information about the buyer's type. Such a state is not represented as a movement in posterior beliefs: it affects the payoff attached to a posterior rather than the set of posteriors the seller can induce \citep{DovalSmolin2026}. The question is therefore whether the mechanism run at a given posterior should be allowed to depend on the realized public state. In screening, the answer is simple: the gain from conditioning on the public state equals the known-type gain, averaged under the prior. Incentive constraints, allocation distortions, and informational rents affect the level of revenue, but not this incremental gain.

\paragraph{Setup.}
The date-\(t\) state is \((S_t,Y_t)\), where \(S_t\) is the seller's posterior belief over the type \(\theta\in\Theta\), and \(Y_t\in\mathcal Y\) is a public payoff state. The public state affects payoffs but carries no information about the buyer's type. Formally, for each \(t\) there is a probability measure \(\nu_t\) on \(\mathcal Y\), independent of the realized posterior, such that
\begin{equation}\tag{Non-posterior}\label{eq:Non-posterior}
\operatorname{Law}(Y_t\mid S_t=\mu)=\nu_t
\quad\text{for all }\mu,
\qquad
\operatorname{Law}(\theta\mid S_t=\mu,Y_t=y)=\mu .
\end{equation}
Thus \(Y_t\) changes the payoff environment without changing the posterior
belief.

For a realized public state \(y\), a direct mechanism allocates with probability \(q\) and charges transfer \(p\), at production cost \(c(y)\). Incentive and participation constraints are imposed pointwise in the realized public state:
\begin{align}
u(\theta,q(\theta,y),y)-p(\theta,y)
&\ge u(\theta,q(\theta',y),y)-p(\theta',y),
&&\forall\,\theta,\theta',y,
\tag{IC}\label{eq:screen-IC}\\
u(\theta,q(\theta,y),y)-p(\theta,y)&\ge 0,
&&\forall\,\theta,y.
\tag{IR}\label{eq:screen-IR}
\end{align}
By \cref{lem:screen-revelation}, the continuation value is a standard screening program. The two regimes differ in whether the schedule is chosen before or after the public state is observed:
\[
\widehat g(\mu)
=
\sup_{t\in\mathbb T}
\int_{\mathcal Y}
\Big(\max_{q,p}\textstyle\int_\Theta [p-c(y)q]\,d\mu\Big)\,\nu_t(dy),
\qquad
g(\mu)
\ge
\sup_{t\in\mathbb T}
\max_{q,p}
\int_{\mathcal Y}\!\int_\Theta [p-c(y)q]\,d\mu\,\nu_t(dy).
\]
Here \(\widehat g\) allows the mechanism to depend on the realized \(y\), while
the displayed lower bound for \(g\) restricts the seller to a single schedule
chosen before \(Y_t\) is observed.\footnote{The inequality allows for the
possibility that a \(y\)-independent mechanism still induces \(y\)-dependent
behavior when \(Y_t\) changes the buyer's ranking over outcomes
(\cref{lem:screen-revelation}).}

\begin{assumptionsub}[Screening regularity]
\label{ass:screen-regularity}
The screening environment satisfies the sufficient conditions in \textup{R2} of
\cref{app:regularity-sufficient}. In particular, \(g\) and \(\widehat g\) are
bounded and upper semicontinuous.
\end{assumptionsub}
\paragraph{Known types.}
In the screening problem, revenue is convex in the posterior. The objective is linear in the belief, and the incentive and participation constraints do not depend on the belief \(\mu\) (see \cref{lem:screen-convex}). Hence the relevant concavified values are evaluated at point-mass posteriors. Thus the comparison between state-contingent and state-independent mechanisms is made type by type (see \cref{cor:screen-known-type}). Splitting the prior into interior posteriors does not change this comparison.

\begin{proposition}[Screening collapse criterion]
\label{prop:screen-collapse}
Suppose \cref{ass:single-shot,ass:screen-regularity} hold. At prior \(S_0\), the
state-contingent and state-independent ex-ante values coincide if and only if
\begin{equation}\label{eq:Screening collapse criterion}
\widehat g(\delta_\theta)=g(\delta_\theta)
\qquad\text{for }S_0\text{-almost every }\theta\in\Theta.
\end{equation}
A sufficient condition is
\begin{equation}
\label{eq:screen-collapse-identity}
\sup_{t\in\mathbb T}
\int_{\mathcal Y}\sup_{q\in[0,1]}
\bigl[u(\theta,q,y)-c(y)q\bigr]\,\nu_t(dy)
\;=\;
\sup_{t\in\mathbb T}\sup_{q\in[0,1]}
\Bigl[
\operatorname*{ess\,inf}_{y\sim\nu_t}u(\theta,q,y)
-
q\!\int_{\mathcal Y}c(y)\,\nu_t(dy)
\Bigr]
\end{equation}
for \(S_0\)-almost every \(\theta\in\Theta\).
\end{proposition}

\begin{proof}
By \cref{prop:value-representation}, the two ex-ante values at \(S_0\) are \((\operatorname{conc}_f\widehat g)(S_0)\) and \((\operatorname{conc}_f g)(S_0)\). Since \(g\le \widehat g\), the identity in \cref{cor:screen-known-type} implies that these values coincide if and only if \eqref{eq:Screening collapse criterion}. For the sufficient condition, fix \(\theta\). By \cref{lem:screen-revelation} evaluated at \(\mu=\delta_\theta\), thus \eqref{eq:screen-collapse-identity} and \(g\le \widehat g\) imply \(g(\delta_\theta)=\widehat g(\delta_\theta)\). If the condition holds for \(S_0\)-almost every \(\theta\), the first part gives equality of the two ex-ante
values.
\end{proof}

\paragraph{Interpreting the criterion.}
The criterion has a direct interpretation. Commitment to a state-independent mechanism is without loss if, for almost every type, the seller would make the same decision even under full information about that type. Thus the comparison is not driven by incentive constraints or informational rents. Those objects affect the level of screening revenue, but the value of conditioning on the public state is evaluated type by type.

Consider the cost-shifter case ($u(\theta,q,y)=\theta q$). For a given date $t$, a seller who observes $Y_t$ before choosing whether to trade obtains $\int_{\mathcal Y}(\theta-c(y))^{+} \, \nu_t(dy)$ at known type $\theta$. If the trade decision cannot depend on $Y_t$, the corresponding payoff is $(\theta-\int_{\mathcal Y}c(y) \, \nu_t(dy))^{+}$. The difference is positive only for types for which the realized cost state changes the efficient trade decision. Types served in every cost state and types served in no cost state contribute nothing to the value of conditioning on $Y_t$. Since the screening problem reduces this comparison to known types, the ex-ante value is obtained by averaging these type-specific differences under the prior. The same logic applies when $u(\theta,q,y)=\theta v(y)q$.

The reduction to known-type comparisons relies on the convexity of screening revenue in the posterior. This feature is not shared by all applications. In the certification problem below \cref{subsec:certification}, the information policy may induce pooling at an endogenous posterior, so the value of conditioning on a public payoff state is evaluated at that posterior rather than at known types. There, information design and public-state flexibility no longer separate.

\subsection{Certification with demand shocks}
\label{subsec:certification}
We apply the criterion to a certification problem. A seller discloses evidence about product quality to obtain approval from a regulator, as in approval-persuasion models \citep{HenryOttaviani,McClellan2022}; conditional on approval, the seller then faces a public demand shock and chooses whether to commit to a price before the shock or adjust the price after it is realized. The demand shock is the non-posterior public state in \cref{sec:collapse-stat}. The question is whether the pass posterior induced by persuasion already absorbs the value of this state, or the state-contingent pricing still raises revenue.

\paragraph{Setup.}
The seller has one unit (\cref{ass:single-shot}). Quality is $\theta\in\Theta$, and $S\in\Delta(\Theta)$ is the public posterior, with mean $m(S):=\sum_{\theta}\theta S(\theta)$. A public demand shock $Y\in\Y$ is realized after approval. It is payoff-relevant for the buyer but carries no further information about quality $\operatorname{Law}(Y\mid S=\mu)=\nu$ as in \eqref{eq:Non-posterior}. The regulator observes the seller's public disclosure and approves the good iff the posterior mean clears a cutoff $m(S)\ge \kappa$ where $\kappa$ is the regulator's reservation belief. Conditional on approval, the seller sells to a buyer whose value is $v(\theta,y)$. For each $y$, $v(\cdot,y)$ is affine and increasing in quality.
Write
\[
v(m,y):=\sum_\theta S(\theta)v(\theta,y),
\qquad
\bar v(m):=\sum_y\nu(y)v(m,y),
\]
where the first expression is evaluated at any posterior with mean $m$. The seller's unit cost is constant and equal to $c$. The prior mean is $m_0:=m(S_0)$. The seller chooses whether the price is fixed before the demand shock or can respond to it after realization. Approval carries a supply obligation (\cref{ass:cert-regularity}): the unit trades in every demand state at the posted price. Under \emph{ex-ante} pricing the price is set before $Y$, so the obligation requires participation in the worst demand state, $p\le\min_{y\in\Y}v(m,y)$; under \emph{state-contingent} pricing the price is set after $Y$, and the binding constraint is $p\le v(m,y)$ state by state. With a single date, the posterior-only and full-history values are
\begin{equation}
\label{eq:cert-values}
g(\mu) =
\ind{m(\mu)\ge\kappa}
\Bigl[\min_{y\in\Y}v(m(\mu),y)-c\Bigr],
\qquad
\widehat g(\mu) =
\ind{m(\mu)\ge\kappa}
\bigl[\bar v(m(\mu))-c\bigr].
\end{equation}
Thus the value of conditioning the price on the demand shock is
\[
\widehat g(\mu)-g(\mu) = 
\ind{m(\mu)\ge\kappa}
\Bigl[\bar v(m(\mu))-\min_{y\in\Y}v(m(\mu),y)\Bigr].
\]

The obstruction is the endogenous location of the gap $\bar v(m)-\min_y v(m,y)$ in \eqref{eq:cert-values}. Persuasion chooses the pass posterior, and the value of conditioning on $Y$ is tested only there. Hence the non-posterior state is redundant only when the induced posterior makes the worst and average demand states coincide; a pure cost shock would make the post-approval value convex and return the comparison to known qualities, as in
\cref{sec:screening-application}.

\begin{assumptionsub}[Regularity]
\label{ass:cert-regularity}
$\Theta$ and $\Y$ are finite; $v(\cdot,y)$ is affine, increasing, and bounded; cost is constant at $c$; the regulator approves above a cutoff in the posterior mean; approval carries a supply obligation, so that the unit trades in every demand state at the posted price; the lowest-quality good is profitable in every demand state, $\min_y v(0,y)\ge c$, so the obligation is never loss-making; and $\conc_f g$ and $\conc_f\widehat g$ are upper semicontinuous at $S_0$.
\end{assumptionsub}

Under \cref{ass:cert-regularity}, the optimal certification test pools substandard beliefs into the lowest passing posterior $m=\kappa$. Thus the pricing comparison is evaluated at the endogenous pass posterior induced by persuasion, rather than at the prior or at known qualities. Call $m^\circ$ the
demand-neutral belief, where $\min_y v(m^\circ,y)=\bar v(m^\circ)$.

\begin{proposition}
\label{prop:cert-collapse}
Suppose \cref{ass:single-shot,ass:cert-regularity} hold and the prior mean is below the standard, $m_0<\kappa$. Ex-ante and state-contingent pricing yield the same revenue if and only if $\kappa=m^{\circ}$. Otherwise state-contingent pricing strictly dominates, with revenue gain
\[
\frac{m_0}{\kappa}\bigl[\bar v(\kappa)-\min_y v(\kappa,y)\bigr],
\]
vanishing as $\kappa\to m^{\circ}$.
\end{proposition}

\begin{proof}
By \cref{prop:collapse-structure} the regimes coincide at $S_0$ if and only if, for every $\varepsilon>0$, a continuous affine $\ell_\lambda$ majorizes $g$ and $\widehat g$ with $\ell_\lambda(S_0)\le(\conc_f g)(S_0)+\varepsilon$; the multiplier $\lambda$ is the persuasion price of \citet{DworczakMartini}. Both values pool to the posterior $\kappa$ (preceding paragraph), so for $m_0<\kappa$, $(\conc_f g)(S_0)=\tfrac{m_0}{\kappa}g(\kappa)$ and $(\conc_f\widehat g)(S_0)=\tfrac{m_0}{\kappa}\widehat g(\kappa)$. Because $\widehat g\ge g$, any affine majorant of $\widehat g$ has value at least $(\conc_f\widehat g)(S_0)$ at $S_0$; such a majorant is admissible---tight against $(\conc_f g)(S_0)$---if and only if $(\conc_f\widehat g)(S_0)=(\conc_f g)(S_0)$, i.e.\ $g(\kappa)=\widehat g(\kappa)$, i.e.\ $\min_y v(\kappa,y)=\bar v(\kappa)$, i.e.\ $\kappa=m^{\circ}$. The deficit $\tfrac{m_0}{\kappa}[\widehat g(\kappa)-g(\kappa)]$ is the stated gain, and \cref{prop:diagnostic} reads the same way: the only pooled belief is $\kappa$, so $\widehat g$ meets rather than enters the gain region precisely when the demand shock is irrelevant there.
\end{proof}

The comparison is therefore made at the pass posterior, not at the prior. Persuasion fixes this posterior at $\kappa$, and the demand shock is redundant only if fixed pricing already attains the state-contingent value there. When $\kappa\neq m^\circ$, the public demand state remains payoff-relevant after approval, so conditioning the price on $Y$ strictly raises revenue.

\paragraph{An illustration.}
Let $\Theta=\{0,1\}$ and $\Y=\{a,b\}$ with $\nu(a)=\nu(b)=\tfrac12$ and $c=0$, and let $v(\theta,a)=0.6+0.2\,\theta$ and $v(\theta,b)=0.2+0.8\,\theta$, so the two demand states rank the buyer's value oppositely in quality. The lowest-quality good is profitable ($\min_y v(0,y)=0.2\ge c$), so the pass posterior is $\kappa$. Then $g(m)=\ind\{m\ge\kappa\}\min(0.6+0.2m,\,0.2+0.8m)$, $\widehat g(m)=\ind\{m\ge\kappa\}(0.4+0.5m)$, and the demand-neutral belief is $m^{\circ}=\tfrac23$. \Cref{fig:cert-gain} takes $\kappa=0.3$: the optimal experiment pools $[0,\kappa]$ into the posterior $\kappa$, and $\widehat g$ exceeds $\conc_f g$ above $\kappa$ except at $m^{\circ}$, so the seller strictly gains from waiting at every prior mean below the standard, the gain increasing in $|\kappa-m^{\circ}|$ in this two-state example.

\begin{figure}[H]
  \centering
  \begin{tikzpicture}
    \begin{axis}[
      width=0.88\textwidth, height=6.2cm, axis lines=left,
      xmin=0, xmax=1, ymin=0, ymax=1.02,
      xtick={0,0.3,0.6667,1}, xticklabels={$0$,$\kappa$,$m^{\circ}$,$1$},
      ytick=\empty, xlabel={Posterior mean $m$}, ylabel={Continuation value},
      label style={font=\small}, ticklabel style={font=\small},
      legend style={font=\scriptsize, at={(0.02,0.98)}, anchor=north west,
                    legend columns=2, draw=none, fill=none},
      legend cell align=left, clip=false]
      \addplot[thick, blue, domain=0.3:0.6667, samples=2] {0.2+0.8*x};
      \addlegendentry{$g$}
      \addplot[thick, blue, domain=0.6667:1, samples=2, forget plot] {0.6+0.2*x};
      \addplot[blue, dashed, forget plot] coordinates {(0.3,0) (0.3,0.44)};
      \addplot[only marks, mark=*, mark size=1.1pt, blue, forget plot] coordinates {(0.3,0.44)};
      \addplot[thick, red, dashed, domain=0.3:1, samples=2] {0.4+0.5*x};
      \addlegendentry{$\widehat g$}
      \addplot[red, dashed, forget plot] coordinates {(0.3,0) (0.3,0.55)};
      \addplot[only marks, mark=*, mark size=1.1pt, red, forget plot] coordinates {(0.3,0.55)};
      \addplot[dotted, thick, gray, domain=0:0.3, samples=2] {(0.44/0.3)*x};
      \addlegendentry{$\conc_f g$}
      \addplot[dotted, thick, gray, domain=0.3:0.6667, samples=2, forget plot] {0.2+0.8*x};
      \addplot[dotted, thick, gray, domain=0.6667:1, samples=2, forget plot] {0.6+0.2*x};
      \addplot[name path=U, draw=none, domain=0.3:1, samples=2, forget plot] {0.4+0.5*x};
      \addplot[name path=L1, draw=none, domain=0.3:0.6667, samples=2, forget plot] {0.2+0.8*x};
      \addplot[name path=L2, draw=none, domain=0.6667:1, samples=2, forget plot] {0.6+0.2*x};
      \addplot[red!10, forget plot] fill between[of=U and L1, soft clip={domain=0.3:0.6667}];
      \addplot[red!10, forget plot] fill between[of=U and L2, soft clip={domain=0.6667:1}];
      \addlegendimage{area legend, fill=red!10, draw=red!40}
      \addlegendentry{$\widehat g-\conc_f g$}
      \addplot[dotted, gray, forget plot] coordinates {(0.2,0) (0.2,0.293)};
      \node[anchor=north, font=\scriptsize] at (axis cs:0.2,0) {$m_0$};
    \end{axis}
  \end{tikzpicture}
  \caption{The illustration with $\kappa=0.3$ and $m^{\circ}=\tfrac23$. Since the
  lowest-quality good is profitable, the optimal experiment pools $[0,\kappa]$
  into the pass posterior $\kappa$. State-contingent pricing $\widehat g$ exceeds
  $\conc_f g$ above $\kappa$ except at $m^{\circ}$ (shaded), so by
  \cref{prop:cert-collapse} the seller gains from waiting unless
  $\kappa=m^{\circ}$.}
  \label{fig:cert-gain}
\end{figure}

\bibliography{reference11}
\appendix
\section{Lemmas and proofs}
\subsection{Lemmas}

\begin{lemma}[Posterior process]\label{lem:posterior-martingale}
For each $t\in\mathbb T$, there exists an $\mathcal F_t$-measurable  posterior kernel $S_t:\Omega\to\Delta(\Theta)$ such that, for every  bounded Borel function $\varphi:\Theta\to\mathbb R$,
\[
\int_\Theta \varphi(\theta')\,S_t(\omega)(d\theta')
=
\mathbb E[\varphi(\theta)\mid\mathcal F_t](\omega)
\qquad\text{a.s.}
\]
In particular:
\begin{enumerate}[label=\textup{(\roman*)}]
    \item $S_t$ is a sufficient statistic for $\theta$ given 
    $\mathcal F_t$: conditioning on $S_t$ and conditioning on 
    $\mathcal F_t$ yield the same conditional distribution of $\theta$.
    \item Since $\mathcal F_t = \sigma(Z_t)$, the Doob--Dynkin lemma 
    implies that $S_t = f_t(Z_t)$ for some Borel measurable function 
    $f_t$; that is, the public history $Z_t$ induces the posterior 
    $S_t$.
    \item $\{S_t\}_{t\in\mathbb T}$ is a bounded 
    $(\mathcal F_t)$-martingale.
\end{enumerate}
\end{lemma}

\begin{proof}

\noindent\textit{Existence of the posterior kernel.}
Since $\Theta$ is a compact Polish space, $\Delta(\Theta)$ equipped with the weak topology is also a compact Polish space, hence a standard Borel space. The map $\theta:\Omega\to\Theta$ is $\mathcal F$-measurable and $\mathcal F_t\subseteq\mathcal F$ is a sub-$\sigma$-algebra. By the standard theorem on regular conditional distributions for Polish-space-valued random variables \citep[Theorem~6.3]{Kallenberg2002}, there exists a Markov kernel $S_t:\Omega\times\mathcal B(\Theta)\to[0,1]$ such that
\begin{enumerate}[label=\textup{(\alph*)}]
    \item
    $\omega\mapsto S_t(\omega)(B)$ is $\mathcal F_t$-measurable for
    every $B\in\mathcal B(\Theta)$, and
    \item
    $S_t(\omega)(\cdot)
    =\mathbb P(\theta\in\cdot\mid\mathcal F_t)(\omega)$
    \quad$\mathbb P$-a.s.
\end{enumerate}
Equivalently, for every bounded Borel $\varphi:\Theta\to\mathbb R$,
\begin{equation}\label{eq:kernel-property}
\int_\Theta\varphi(\theta')\,S_t(\omega)(d\theta')
=\mathbb E[\varphi(\theta)\mid\mathcal F_t](\omega)
\qquad\text{a.s.}
\end{equation}

\medskip
\noindent\textit{Part \textup{(i)}: Sufficiency.}
Fix any bounded Borel $\varphi:\Theta\to\mathbb R$ and define
$F:\Delta(\Theta)\to\mathbb R$ by
$F(\mu):=\int_\Theta\varphi(\theta')\,\mu(d\theta')$.
By the functional monotone class theorem, $F$ is Borel measurable on $\Delta(\Theta)$: it holds for continuous $\varphi$ by the definition of the weak topology, and extends to all bounded Borel $\varphi$ by dominated convergence.\footnote{Explicitly: every bounded Borel function is the pointwise limit of a uniformly bounded sequence of continuous functions (by Lusin's theorem and dominated convergence), so $F(\mu)=\lim_n\int\varphi_n\,d\mu$ is a pointwise limit of Borel measurable functions, hence Borel measurable.} Therefore $\omega\mapsto F(S_t(\omega))$ is
$\sigma(S_t)$-measurable.
By \eqref{eq:kernel-property},
$\mathbb E[\varphi(\theta)\mid\mathcal F_t](\omega)=F(S_t(\omega))$
a.s., so $\mathbb E[\varphi(\theta)\mid\mathcal F_t]$ is
$\sigma(S_t)$-measurable.
For any $B\in\sigma(S_t)\subseteq\mathcal F_t$,
\[
\mathbb E\bigl[\varphi(\theta)\mathbf 1_B\bigr]
=\mathbb E\bigl[\mathbb E[\varphi(\theta)\mid\mathcal F_t]
\mathbf 1_B\bigr]
=\mathbb E\bigl[F(S_t)\mathbf 1_B\bigr].
\]
Since $F(S_t)$ is $\sigma(S_t)$-measurable, uniqueness of conditional
expectation gives
$\mathbb E[\varphi(\theta)\mid\mathcal F_t]
=\mathbb E[\varphi(\theta)\mid S_t]$ a.s.
As $\varphi$ was arbitrary, conditioning on $S_t$ and conditioning on
$\mathcal F_t$ yield the same conditional distribution of $\theta$.

\medskip
\noindent\textit{Part \textup{(ii)}: Induction by $Z_t$.}
Since $\mathcal F_t=\sigma(Z_t)$ and $S_t$ is $\sigma(Z_t)$-measurable with values in the standard Borel space $\Delta(\Theta)$, the
Doob--Dynkin lemma
\citep[Lemma~1.14]{Kallenberg2002}
yields a Borel measurable function
$f_t:E_t\to\Delta(\Theta)$ such that
\[
S_t(\omega)=f_t(Z_t(\omega))
\qquad\mathbb P\text{-a.s.}
\]
Hence $Z_t$ induces the posterior $S_t$ through the measurable map $f_t$.

\medskip
\noindent\textit{Part \textup{(iii)}: Martingale property.}
Fix $s\le t$ and any bounded Borel $\varphi:\Theta\to\mathbb R$.
Then
\[
\mathbb E\!\left[
\int_\Theta\varphi(\theta')\,S_t(\omega)(d\theta')
\,\Big|\,\mathcal F_s
\right]
\stackrel{\eqref{eq:kernel-property}}{=}
\mathbb E\bigl[
\mathbb E[\varphi(\theta)\mid\mathcal F_t]
\mid\mathcal F_s
\bigr]
\stackrel{\text{tower}}{=}
\mathbb E[\varphi(\theta)\mid\mathcal F_s]
\stackrel{\eqref{eq:kernel-property}}{=}
\int_\Theta\varphi(\theta')\,S_s(\omega)(d\theta')
\quad\text{a.s.},
\]
where the second equality uses the tower property of conditional expectations and $\mathcal F_s\subseteq\mathcal F_t$.
Since $\varphi$ was arbitrary, $\mathbb E[S_t\mid\mathcal F_s]=S_s$ a.s.\ in $\Delta(\Theta)$. Boundedness holds because $S_t(\omega)$ is a probability measure:
for every bounded $\varphi$,
$\bigl|\int_\Theta\varphi\,dS_t\bigr|\le\|\varphi\|_\infty<\infty$.
\end{proof}

\begin{lemma}[Payoff selections]\label{lem:selection-exists}
Suppose \cref{ass:analytic-equilibrium} holds. For any continuation mechanism rule $\Gamma_t$, the correspondence  $\mathcal R_t^{\Gamma_t}$ admits a universally measurable selection,  and every such selection is bounded above by a common finite constant.
\end{lemma}

\begin{proof}
By assumption, $\mathcal R_t^{\Gamma_t}$ has nonempty values, analytic  graph in $\Omega\times\mathbb R$, and is uniformly bounded above by some  $\bar r<\infty$. Since $\Omega$ and $\mathbb R$ are standard Borel spaces  and the projection of the graph onto $\Omega$ is all of $\Omega$, the Jankov--von Neumann selection theorem yields a universally measurable map \citep[Theorem~18.1]{Kechris}
$\rho_t:\Omega\to\mathbb R$ with 
$\rho_t(\omega)\in\mathcal R_t^{\Gamma_t}(\omega)$ for all $\omega$, 
hence $\rho_t\in\Sel(\mathcal R_t^{\Gamma_t})$.  Uniform boundedness gives $\rho_t(\omega)\le\bar r$ a.s.\ for every such selection.
\end{proof}

\begin{lemma}[Measurable equilibrium lifting]
\label{lem:measurable-lifting}
Suppose \cref{ass:analytic-equilibrium} holds. For every continuation mechanism rule $\Gamma_t$ and every payoff selection
$\rho_t\in\Sel(\mathcal R_t^{\Gamma_t})$, there exists a universally measurable map
\[
b_t:\Omega\to\mathcal B_t
\]
such that, $\mathbb P$-a.e.,
\[
b_t(\omega)\in
\mathcal E_t\bigl(Z_t(\omega),\Gamma_t(\omega)\bigr)
\quad\text{and}\quad
\rho_t(\omega)
=
\pi_t\bigl(\omega,Z_t(\omega),\Gamma_t(\omega),b_t(\omega)\bigr).
\]
\end{lemma}

\begin{proof}
Fix a continuation mechanism rule $\Gamma_t$ and a payoff selection
$\rho_t\in\Sel(\mathcal R_t^{\Gamma_t})$.

Define
\[
C_t
:=
\Bigl\{
(\omega,r,b)\in\Omega\times\mathbb R\times\mathcal B_t:
b\in\mathcal E_t(Z_t(\omega),\Gamma_t(\omega)),
\quad
r=\pi_t(\omega,Z_t(\omega),\Gamma_t(\omega),b)
\Bigr\}.
\]
The set \(C_t\) is analytic. Indeed, the equilibrium condition is the measurable pullback of the analytic graph \(\operatorname{Gr}(\mathcal E_t)\), and the payoff equality is Borel because \(\pi_t\) is jointly Borel measurable. Hence, \(C_t\) is the
intersection of an analytic set and a Borel set.

By the Jankov--von Neumann selection theorem
\citep[Theorem~18.1]{Kechris}, the projection of \(C_t\) on \(\Omega\times\mathbb R\) admits a universally measurable uniformization:
there is a universally measurable map \(j_t\) such that
\[
(\omega,r,j_t(\omega,r))\in C_t
\]
whenever \((\omega,r)\) belongs to that projection.

Since \(\rho_t\in\Sel(\mathcal R_t^{\Gamma_t})\), we have
\[
(\omega,\rho_t(\omega))
\in
\operatorname{proj}_{\Omega\times\mathbb R}C_t
\qquad \mathbb P\text{-a.e.}
\]
The map \(\omega\mapsto(\omega,\rho_t(\omega))\) is universally measurable. Replacing \(j_t\), under the image of \(\mathbb P\) by this map, with a Borel version if necessary
\citep[Chapter~7]{BertsekasShreve}, define
\[
b_t(\omega):=j_t(\omega,\rho_t(\omega)).
\]
Then \(b_t\) is universally measurable, and
\[
(\omega,\rho_t(\omega),b_t(\omega))\in C_t
\qquad \mathbb P\text{-a.e.}
\]
By the definition of \(C_t\), this means
\[
b_t(\omega)\in
\mathcal E_t(Z_t(\omega),\Gamma_t(\omega))
\]
and
\[
\rho_t(\omega)
=
\pi_t(\omega,Z_t(\omega),\Gamma_t(\omega),b_t(\omega))
\]
\(\mathbb P\)-a.e.
\end{proof}

\begin{lemma}[Pasting of continuation equilibria]
\label{lem:sequential-pasting}
Let $\pi=\sum_{k=1}^K\alpha_k\delta_{\mu_k}$ be a finite-support posterior distribution induced by a public experiment at date \(0\). For each \(k\), let \(t_k\in\mathbb T\), and let
\((\Gamma_k,\rho_k)\) be an \(\varepsilon\)-optimal selector for \(U_{t_k}^{\mathcal F_{t_k}}(\mu_k)\).
Under \Cref{ass:analytic-equilibrium,ass:single-shot}, there exists a strategy--belief
profile of the resulting dynamic game that is sequentially rational and whose seller payoff is at least $\sum_{k=1}^K
\alpha_k U_{t_k}^{\mathcal F_{t_k}}(\mu_k)-\varepsilon$.
\end{lemma}

\begin{proof}
Let \(J\in\{1,\dots,K\}\) denote the public signal generated by the date-0 experiment, with
\[
\mathbb P(J=k)=\alpha_k,
\qquad
\mathbb P(\theta\in\cdot\mid J=k)=\mu_k .
\]

Fix \(k\). Since \(\rho_k\in \Sel\bigl(\mathcal R_{t_k}^{\Gamma_k}\bigr)\), \cref{lem:measurable-lifting} yields a universally measurable continuation-equilibrium selector \(b_k\) such that
\[
b_k(\omega)\in
\mathcal E_{t_k}\bigl(Z_{t_k}(\omega),\Gamma_k(\omega)\bigr)
\quad\text{and}\quad
\rho_k(\omega)
=
\pi_{t_k}\bigl(\omega,Z_{t_k}(\omega),\Gamma_k(\omega),b_k(\omega)\bigr)
\qquad \mathbb P\text{-a.e.}
\]

Construct a strategy--belief profile as follows. On branch \(\{J=k\}\), assign the date-\(t_k\) continuation rule \(\Gamma_k\) and the continuation strategy--belief selector \(b_k\). At public histories off the support of the date-0 experiment, assign an arbitrary continuation equilibrium from the corresponding nonempty continuation-equilibrium set.

By construction, after every public history reached on branch \(\{J=k\}\), the prescribed continuation play belongs to \(\mathcal E_{t_k}\bigl(Z_{t_k},\Gamma_k\bigr)\). Off the support of the experiment, continuation play is also chosen from the relevant continuation-equilibrium set. Hence the pasted strategy--belief profile is sequentially rational.

Conditional on branch \(\{J=k\}\), the seller's payoff is
\(\int_\Omega\rho_k(\omega)\,\mathbb P_{t_k}^{\mu_k}(d\omega)\). Since \((\Gamma_k,\rho_k)\) is \(\varepsilon\)-optimal for \(U_{t_k}^{\mathcal F_{t_k}}(\mu_k)\),
\[
\int_\Omega\rho_k(\omega)\,\mathbb P_{t_k}^{\mu_k}(d\omega)
\ge
U_{t_k}^{\mathcal F_{t_k}}(\mu_k)-\varepsilon .
\]
Averaging over \(J\), the seller's ex-ante payoff is at least
\[
\sum_{k=1}^K\alpha_k\bigl(U_{t_k}^{\mathcal F_{t_k}}(\mu_k)-\varepsilon\bigr)
=\sum_{k=1}^K\alpha_k U_{t_k}^{\mathcal F_{t_k}}(\mu_k)-\varepsilon ,
\]
because \(\sum_{k=1}^K\alpha_k=1\). This proves the claim.
\end{proof}

\begin{lemma}[Upper bound]\label{lem:upper-bound}
Suppose \cref{ass:single-shot,ass:regularity} hold. For every admissible mechanism calendar \(\mathcal M\) and every sequentially rational Bayesian equilibrium of the induced dynamic game, the seller's ex-ante equilibrium
payoff satisfies
\[
R(\mathcal M)
\;\le\;
\Big(\operatorname{conc}_f\big[\sup_{t\in\mathbb T}U_t^{\mathcal F_t}\big]\Big)(S_0).
\]
Consequently,
\[
\sup_{\mathcal M}R(\mathcal M)
\;\le\;
\Big(\operatorname{conc}_f\big[\sup_{t\in\mathbb T}U_t^{\mathcal F_t}\big]\Big)(S_0).
\]
\end{lemma}

\begin{proof}
Fix an admissible mechanism calendar \(\mathcal M\) and a sequentially rational Bayesian equilibrium with strategy--belief profile \(b\). Let \(\xi\) denote the seller's realized equilibrium payoff. Since date \(0\) is the public disclosure stage and no allocation or transfer is made before the induced date-\(1\) continuation problem, the seller's realized payoff is the payoff of that continuation problem.

By sequential rationality,
\[
b(\omega)\in
\mathcal E_1\bigl(Z_1(\omega),\Gamma_1(\omega)\bigr)
\qquad \mathbb P\text{-a.s.}
\]
Define
\[
\rho_1(\omega)
:=
\pi_1\!\bigl(
\omega,Z_1(\omega),\Gamma_1(\omega),b(\omega)
\bigr).
\]
Because \(\pi_1\) is the seller's payoff in the date-\(1\) continuation problem,
\[
\xi=\rho_1
\qquad \mathbb P\text{-a.s.}
\]
Moreover,
\[
\rho_1\in\Sel\!\bigl(\mathcal R_1^{\Gamma_1}\bigr).
\]
Since \(\Gamma_1\) is \(\mathcal F_1\)-measurable, \(\rho_1\) is admissible for
\(U_1^{\mathcal F_1}\). Hence, by the definition of \(U_1^{\mathcal F_1}\),
\[
\int_\Omega \rho_1(\omega)\,\mathbb P_1^\mu(d\omega)
\le
U_1^{\mathcal F_1}(\mu)
\qquad\text{for every }\mu\in\Delta(\Theta).
\]
Equivalently,
\[
\mathbb E[\rho_1\mid S_1]
\le
U_1^{\mathcal F_1}(S_1)
\qquad \mathbb P\text{-a.s.}
\]

Let
\[
\widehat g(\mu):=\sup_{t\in\mathbb T}U_t^{\mathcal F_t}(\mu),
\qquad
\phi(\mu):=
\Big(\operatorname{conc}_f \widehat g\Big)(\mu).
\]
The function \(\phi\) is concave by construction of the finite concavification,
and \(\widehat g\le\phi\). Since \(U_1^{\mathcal F_1}\le\widehat g\), we have
\[
\mathbb E[\xi]
=
\mathbb E[\rho_1]
=
\mathbb E\!\left[\mathbb E[\rho_1\mid S_1]\right]
\le
\mathbb E\!\left[U_1^{\mathcal F_1}(S_1)\right]
\le
\mathbb E[\phi(S_1)].
\]

By \cref{ass:regularity}, \(\phi\) is upper semicontinuous at \(S_0\). Applying the supporting hyperplane theorem \citep[Section~5.12]{AliprantisBorder} to the hypograph of the concave function \(\phi\), for every \(\varepsilon>0\) there exists
\(\ell_\varepsilon\in\mathrm{Aff}_c(\Delta(\Theta))\) such that
\[
\ell_\varepsilon\ge \phi
\quad\text{on }\Delta(\Theta),
\qquad
\ell_\varepsilon(S_0)\le \phi(S_0)+\varepsilon .
\]
By \cref{lem:posterior-martingale}, \(S_t\) is a martingale, so
\[
\mathbb E[S_1]=S_0.
\]
Therefore
\[
\mathbb E[\phi(S_1)]
\le
\mathbb E[\ell_\varepsilon(S_1)]
=
\ell_\varepsilon\!\bigl(\mathbb E[S_1]\bigr)
=
\ell_\varepsilon(S_0)
\le
\phi(S_0)+\varepsilon .
\]
Letting \(\varepsilon\downarrow0\) gives
\[
\mathbb E[\xi]
\le
\phi(S_0)
=
\Big(\operatorname{conc}_f
\big[\sup_{t\in\mathbb T}U_t^{\mathcal F_t}\big]\Big)(S_0).
\]

Since the sequentially rational Bayesian equilibrium was arbitrary, \(R(\mathcal M)\le\phi(S_0)\). Taking the supremum over admissible mechanism calendars completes the proof.
\end{proof}

\begin{lemma}[Finite concavification]
\label{lem:concf-conc}
Let \(q:\Delta(\Theta)\to\mathbb R\) be bounded above. Suppose that \(\operatorname{conc}_f q\) is real-valued on \(\Delta(\Theta)\) and upper semicontinuous at \(S_0\). Then
\[
(\operatorname{conc}_f q)(S_0)
=
(\operatorname{conc}q)(S_0).
\]
\end{lemma}

\begin{proof}
Let
\[
\phi:=\operatorname{conc}_f q.
\]
If \(\ell\in\mathrm{Aff}_c(\Delta(\Theta))\) satisfies \(\ell\ge q\), then for every finite Bayes-plausible distribution
\[
\sum_{k=1}^K\alpha_k\delta_{\mu_k}
\quad\text{with}\quad
\sum_{k=1}^K\alpha_k\mu_k=\mu,
\]
affineness gives
\[
\sum_{k=1}^K\alpha_k q(\mu_k)
\le
\sum_{k=1}^K\alpha_k\ell(\mu_k)
=
\ell(\mu).
\]
Taking the supremum over all such distributions yields
\[
\phi(\mu)\le \ell(\mu)
\qquad
\forall\,\mu\in\Delta(\Theta).
\]
Taking the infimum over all continuous affine majorants \(\ell\) of \(q\),
\[
\phi(S_0)\le(\operatorname{conc}q)(S_0).
\]

For the reverse inequality, note that \(q\le\phi\) and that \(\phi\) is concave. Since \(\phi\) is upper semicontinuous at \(S_0\), the supporting hyperplane theorem applied to the hypograph of \(\phi\) \citep[Section~5.12]{AliprantisBorder} gives, for every \(\varepsilon>0\), some
\[
\ell_\varepsilon\in\mathrm{Aff}_c(\Delta(\Theta))
\]
such that
\[
\phi\le\ell_\varepsilon
\quad\text{on }\Delta(\Theta),
\qquad
\ell_\varepsilon(S_0)\le\phi(S_0)+\varepsilon.
\]
Since \(q\le\phi\), \(\ell_\varepsilon\) is a continuous affine majorant of
\(q\). Hence
\[
(\operatorname{conc}q)(S_0)
\le
\ell_\varepsilon(S_0)
\le
\phi(S_0)+\varepsilon.
\]
Letting \(\varepsilon\downarrow0\) gives
\[
(\operatorname{conc}q)(S_0)\le\phi(S_0).
\]
Therefore
\[
(\operatorname{conc}_f q)(S_0)
=
(\operatorname{conc}q)(S_0).
\]
\end{proof}

\begin{lemma}[Optimal-stopping representation]\label{lem:stopping-form}
Under \cref{ass:single-shot}, the continuation value is the Snell envelope of the truncation values along the posterior process: for every public state $\{\mathcal G_s\}$, every $t$, and
every $\mu$,
\begin{equation}
U_t^{\mathcal G_t}(\mu)=\max\Bigl\{V_t^{\mathcal G_t}(\mu),\
\int_\Omega U_{t+1}^{\mathcal G_{t+1}}\bigl(S_{t+1}(\omega)\bigr)\,\mathbb P_t^\mu(d\omega)\Bigr\}
=\sup_{\tau\ge t}\ \mathbb E^\mu\!\bigl[V_\tau^{\mathcal G_\tau}(S_\tau)\mid\mathcal F_t\bigr].
\end{equation}
\end{lemma}

\begin{proof}
By \cref{ass:single-shot} the resource is allocated at most once. At date $t$ the seller
either allocates, which terminates the problem with within-period payoff $V_t^{\mathcal G_t}(\mu)$, or defers, reaching the date-$(t+1)$ continuation at the updated posterior; the larger of the two is the first equality. Iterating, an optimal policy is a stopping rule for the posterior process, and $U_t^{\mathcal G_t}(S_t)$ is the least supermartingale dominating $\{V_s^{\mathcal G_s}(S_s)\}_{s\ge t}$, its Snell envelope, which is the second equality \citep{peskir2006optimal}.
\end{proof}

\begin{lemma}[Affine certificates see only the truncation value]\label{lem:cap-equiv}
For continuous $\lambda\colon\Theta\to\mathbb R$ and the affine shadow value
$\ell(\mu)=\int_\Theta\lambda(\theta)\,\mu(d\theta)$,
\begin{equation}\label{eq:cap-equiv}
U_t^{\mathcal G_t}(\mu)\le\ell(\mu)\ \text{for all }t,\mu
\quad\Longleftrightarrow\quad
V_t^{\mathcal G_t}(\mu)\le\ell(\mu)\ \text{for all }t,\mu.
\end{equation}
Consequently
$\bigl(\operatorname{conc}_f[\sup_{t}U_t^{\sigma(S_t)}]\bigr)(S_0)
=\bigl(\operatorname{conc}_f[\sup_{t}V_t^{\sigma(S_t)}]\bigr)(S_0)$.
\end{lemma}

\begin{proof}
By \cref{lem:posterior-martingale} the posterior is a martingale, and $\ell$ is continuous and affine, so $\ell(S_t)$ is a bounded martingale; optional sampling gives $\mathbb E^\mu[\ell(S_\tau)\mid\mathcal F_t]=\ell(\mu)$ for every stopping time $\tau\ge t$. If $\ell\ge V_s^{\mathcal G_s}$ for every $s$, then by \cref{lem:stopping-form}, for each such $\tau$,
$\mathbb E^\mu[V_\tau^{\mathcal G_\tau}(S_\tau)\mid\mathcal F_t]
\le\mathbb E^\mu[\ell(S_\tau)\mid\mathcal F_t]=\ell(\mu)$, so
$U_t^{\mathcal G_t}(\mu)=\sup_{\tau\ge t}\mathbb E^\mu[V_\tau^{\mathcal G_\tau}(S_\tau)\mid\mathcal F_t]\le\ell(\mu)$;
the reverse implication is immediate from $U_t^{\mathcal G_t}\ge V_t^{\mathcal G_t}$. The case
$\sigma(S_t)$ is identical.

It remains to prove the concavification identity. Write
$q_U:=\sup_t U_t^{\sigma(S_t)}$ and $q_V:=\sup_t V_t^{\sigma(S_t)}$. Taking $\tau=t$ in
\cref{lem:stopping-form} gives $U_t^{\sigma(S_t)}\ge V_t^{\sigma(S_t)}$, hence $q_V\le q_U$.
Conversely, fix $\mu\in\Delta(\Theta)$ and $t\in\mathbb T$. By \cref{lem:stopping-form}, for every stopping time $\tau\ge t$ the conditional law of $S_\tau$ given $S_t=\mu$ is a finitely supported, Bayes-plausible distribution with barycenter $\mu$, say $\sum_j\gamma_j\delta_{\nu_j}$; since $V_\tau^{\sigma(S_\tau)}\le q_V$,
\[
\mathbb E^\mu\!\bigl[V_\tau^{\sigma(S_\tau)}(S_\tau)\mid\mathcal F_t\bigr]
=\sum_j\gamma_j\,V_\tau^{\sigma(S_\tau)}(\nu_j)
\le\sum_j\gamma_j\,q_V(\nu_j)
\le(\operatorname{conc}_f q_V)(\mu).
\]
Taking the supremum over $\tau\ge t$ and over $t$ yields $q_U(\mu)\le(\operatorname{conc}_f q_V)(\mu)$
for every $\mu$. Now $\operatorname{conc}_f$ is monotone and idempotent, because the composition of two finite Bayes-plausible splits is again a finite Bayes-plausible split. Applying
$\operatorname{conc}_f$ to the chain $q_V\le q_U\le\operatorname{conc}_f q_V$,
\[
\operatorname{conc}_f q_V\ \le\ \operatorname{conc}_f q_U\ \le\
\operatorname{conc}_f(\operatorname{conc}_f q_V)\ =\ \operatorname{conc}_f q_V,
\]
so $\operatorname{conc}_f q_U=\operatorname{conc}_f q_V$ on $\Delta(\Theta)$; in particular the two agree
at $S_0$.
\end{proof}

\begin{lemma}[Revelation reduction in screening]
\label{lem:screen-revelation}
Fix \(t\in\mathbb T\) and \(\mu\in\Delta(\Theta)\).

\smallskip
\noindent\textup{(i) Full-history.}
If the continuation mechanism may condition on
\(\mathcal F_t=\sigma(S_t,Y_t)\), then
\[
V_t^{\mathcal F_t}(\mu)
=
\sup_{q,p}
\int_{\mathcal Y}\!\!\int_\Theta
\bigl[p(\theta,y)-c(y)q(\theta,y)\bigr]\,
\mu(d\theta)\nu_t(dy),
\]
where the supremum is over measurable
\(q:\Theta\times\mathcal Y\to[0,1]\) and
\(p:\Theta\times\mathcal Y\to\mathbb R\) satisfying
\[
u(\theta,q(\theta,y),y)-p(\theta,y)
\ge
u(\theta,q(\theta',y),y)-p(\theta',y),
\qquad
\forall \theta,\theta',y,
\tag{IC}
\]
and
\[
u(\theta,q(\theta,y),y)-p(\theta,y)\ge0,
\qquad
\forall \theta,y.
\tag{IR}
\]

\smallskip
\noindent\textup{(ii) Posterior-only lower bound.}
If the continuation mechanism is restricted to be
\(\sigma(S_t)\)-measurable, then
\[
V_t^{\sigma(S_t)}(\mu)
\ge
\sup_{q,p}
\int_{\mathcal Y}\!\!\int_\Theta
\bigl[p(\theta)-c(y)q(\theta)\bigr]\,
\mu(d\theta)\nu_t(dy),
\]
where the supremum is over measurable
\(q:\Theta\to[0,1]\) and \(p:\Theta\to\mathbb R\) satisfying
\[
u(\theta,q(\theta),y)-p(\theta)
\ge
u(\theta,q(\theta'),y)-p(\theta'),
\qquad
\forall \theta,\theta',y,
\tag{IC-S}
\]
and
\[
u(\theta,q(\theta),y)-p(\theta)\ge0,
\qquad
\forall \theta,y.
\tag{IR-S}
\]

\end{lemma}

\begin{proof}
By \cref{prop:criterion transform}, the collapse criterion and diagnostic are determined by the truncation values $V_t^{\bullet}$ rather than by the continuation values $U_t^{\bullet}$; it therefore suffices to characterise the static program $V_t$, which we now do. Recall the truncation of the date-$t$ subgame: every public continuation history reaching the start of date $t+1$ is replaced by the absorbing no-sale node (seller payoff $0$, buyer reservation $0$), with truncation-equilibrium set $\mathcal E_t^{\circ}(z,h)$ and truncation payoff correspondence $\mathcal V_t^{\Gamma_t}$. Because the truncated date-$t$ game is itself a continuation game in the sense of \cref{subsec:reduced-form}, \cref{ass:analytic-equilibrium} applies to it, and \cref{lem:selection-exists,lem:measurable-lifting} hold verbatim with $(\mathcal E_t^{\circ},\mathcal V_t^{\Gamma_t})$ in place of $(\mathcal E_t,\mathcal R_t^{\Gamma_t})$.

Let $\nu_t\in\Delta(\mathcal Y)$ be the conditional law of $Y_t$ given $S_t=\mu$. By the non-posterior property \eqref{eq:Non-posterior}, for every bounded measurable $f:\Theta\times\mathcal Y\to\mathbb R$,
\begin{equation}\label{eq:NPexpect-V}
\int_\Omega f\bigl(\theta(\omega),Y_t(\omega)\bigr)\,\mathbb P_t^\mu(d\omega)
=\int_{\mathcal Y}\!\!\int_\Theta f(\theta,y)\,\mu(d\theta)\nu_t(dy).
\end{equation}

\medskip
\noindent\textbf{Part (i): Full-history.}

\emph{The displayed supremum is at most $V_t^{\mathcal F_t}(\mu)$.}
Let $(q,p)$ be measurable maps satisfying \eqref{eq:screen-IC}--\eqref{eq:screen-IR}.
View $(q,p)$ as the direct-revelation mechanism $\Gamma_t\in\mathcal H_t$ that, at public state $(S_t,Y_t)=(\mu,y)$, prescribes outcome $(q(\hat\theta,y),p(\hat\theta,y))$ to report $\hat\theta\in\Theta$ and the null outcome on non-participation; since the outcome rule depends on $y$, $\Gamma_t$ is $\mathcal F_t$-measurable. In the truncated game date $t$ is the only payoff-relevant date: by \eqref{eq:screen-IC} truth-telling $\hat\theta=\theta$ is sequentially rational at every realised $y$, and by \eqref{eq:screen-IR} it is individually rational against the reservation $0$, so truth-telling constitutes a truncation equilibrium $b\in\mathcal E_t^{\circ}(Z_t,\Gamma_t)$ with realised seller payoff
\[
\rho(\omega):=\pi_t(\omega,Z_t,\Gamma_t,b)
=p\bigl(\theta(\omega),Y_t(\omega)\bigr)
-c\bigl(Y_t(\omega)\bigr)q\bigl(\theta(\omega),Y_t(\omega)\bigr)
\in\Sel\bigl(\mathcal V_t^{\Gamma_t}\bigr).
\]
By \eqref{eq:NPexpect-V},
\[
\int_\Omega \rho\,\mathbb P_t^\mu(d\omega)
=\int_{\mathcal Y}\!\!\int_\Theta\bigl[p(\theta,y)-c(y)q(\theta,y)\bigr]\,
\mu(d\theta)\nu_t(dy)\le V_t^{\mathcal F_t}(\mu).
\]
Taking the supremum over admissible $(q,p)$ yields the bound.

\emph{The displayed supremum is at least $V_t^{\mathcal F_t}(\mu)$.}
Fix any $\mathcal F_t$-measurable $\Gamma_t$ and any
$\rho\in\Sel(\mathcal V_t^{\Gamma_t})$. By the truncation form of
\cref{lem:measurable-lifting} there is $b\in\mathcal E_t^{\circ}(Z_t,\Gamma_t)$
with $\rho(\omega)=\pi_t(\omega,Z_t,\Gamma_t,b)$ $\mathbb P$-a.s. The equilibrium
outcome induced by $(\Gamma_t,b)$ is summarised by measurable reduced-form maps
$q_b:\Theta\times\mathcal Y\to[0,1]$ and $p_b:\Theta\times\mathcal Y\to\mathbb R$,
so that
\[
\rho(\omega)=p_b\bigl(\theta(\omega),Y_t(\omega)\bigr)
-c\bigl(Y_t(\omega)\bigr)q_b\bigl(\theta(\omega),Y_t(\omega)\bigr)
\qquad\mathbb P\text{-a.s.}
\]
At each realised $y$ a type-$\theta$ buyer may imitate the truncation-equilibrium behaviour of any type $\theta'$ at the same $y$; sequential rationality therefore gives \eqref{eq:screen-IC} for $(q_b,p_b)$, and participation against the reservation $0$ gives \eqref{eq:screen-IR}. Hence $(q_b,p_b)$ is feasible for the displayed full-history program, and by \eqref{eq:NPexpect-V},
\[
\int_\Omega \rho\,\mathbb P_t^\mu(d\omega)
=\int_{\mathcal Y}\!\!\int_\Theta\bigl[p_b(\theta,y)-c(y)q_b(\theta,y)\bigr]\,
\mu(d\theta)\nu_t(dy),
\]
which is at most the displayed supremum. Taking the supremum over all admissible $(\Gamma_t,\rho)$ gives $V_t^{\mathcal F_t}(\mu)$ at most the displayed supremum. Combining the two inequalities yields equality.

\medskip
\noindent\textbf{Part (ii): Posterior-only lower bound.}
Let $q:\Theta\to[0,1]$ and $p:\Theta\to\mathbb R$ satisfy \textup{(IC-S)} and \textup{(IR-S)}. The direct-revelation mechanism with $y$-independent outcome $(q(\hat\theta),p(\hat\theta))$ is $\sigma(S_t)$-measurable. By \textup{(IC-S)} truth-telling is incentive compatible at every realised $y$, and by \textup{(IR-S)} it is individually rational against the reservation $0$, so it is a truncation equilibrium with realised payoff
\[
\rho(\omega):=p\bigl(\theta(\omega)\bigr)
-c\bigl(Y_t(\omega)\bigr)q\bigl(\theta(\omega)\bigr)
\in\Sel\bigl(\mathcal V_t^{\Gamma_t}\bigr).
\]
By \eqref{eq:NPexpect-V},
\[
\int_\Omega \rho\,\mathbb P_t^\mu(d\omega)
=\int_{\mathcal Y}\!\!\int_\Theta\bigl[p(\theta)-c(y)q(\theta)\bigr]\,
\mu(d\theta)\nu_t(dy)\le V_t^{\sigma(S_t)}(\mu).
\]
Taking the supremum over admissible $(q,p)$ yields the stated bound.

\smallskip
\noindent\emph{The reverse inequality need not hold.}
Applying the construction of part \textup{(i)} to a $\sigma(S_t)$-measurable
$\Gamma_t$ yields reduced-form maps $(q_b,p_b)$ in which $y$ may enter through the
buyer's equilibrium behaviour: although the outcome rule of $\Gamma_t$ is
$y$-independent, her preferences $u(\theta,q,y)-p$ may depend on $y$, so her
optimal report at type $\theta$ may vary with the realised $y$. The induced $(q_b,p_b)$ is then generally $y$-dependent and falls outside the $y$-independent class on the right-hand side. Thus the displayed lower bound corresponds to the stronger requirement that a single $y$-independent schedule implement truthful behaviour at every realisation of $y$, which exceeds $\sigma(S_t)$-measurability of the seller's continuation mechanism.
\end{proof}

\begin{lemma}[Convexity]
\label{lem:screen-convex}
Suppose IC and IR are imposed pointwise in \((\theta,y)\) as in
\eqref{eq:screen-IC}--\eqref{eq:screen-IR}. Then for every \(t\in\mathbb T\), $\mu\mapsto U_t^{\sigma(S_t)}(\mu)$ and $\mu\mapsto U_t^{\mathcal F_t}(\mu)$ are convex on \(\Delta(\Theta)\). Consequently,
\[
g(\mu):=\sup_t U_t^{\sigma(S_t)}(\mu),
\qquad
\widehat g(\mu):=\sup_t U_t^{\mathcal F_t}(\mu)
\]
are convex.
\end{lemma}

\begin{proof}
Fix \(t\). We first prove convexity of
\(U_t^{\sigma(S_t)}\). Consider any \(\sigma(S_t)\)-admissible continuation mechanism and any continuation-equilibrium payoff selection generated by it. In the screening environment, the induced seller payoff can be written as a measurable payoff kernel
\[
r(\theta,y)
=
p_b(\theta,y)-c(y)q_b(\theta,y),
\]
where \((q_b,p_b)\) is the reduced-form allocation-transfer outcome generated by the mechanism together with the buyer's equilibrium behavior. For posterior-only mechanisms, the seller's mechanism itself cannot depend on \(y\), but the induced reduced-form outcome may depend on \(y\) through the buyer's equilibrium behavior.

The pointwise IC and IR restrictions determine whether this reduced-form outcome is implementable. These restrictions may depend on \(y\), but they do not depend on the posterior weights \(\mu\). Thus, once such an equilibrium payoff kernel \(r\) is fixed, changing \(\mu\) changes only the weights assigned to types in the seller's expected payoff:
\[
\int_{\mathcal Y}\!\int_\Theta r(\theta,y)\,\mu(d\theta)\nu_t(dy).
\]
This expression is affine in \(\mu\). Since
\(U_t^{\sigma(S_t)}(\mu)\) is the supremum over all
\(\sigma(S_t)\)-admissible continuation-equilibrium payoff kernels, it
is the supremum of affine functions of \(\mu\). Hence \(U_t^{\sigma(S_t)}\) is convex.

We next prove convexity of \(U_t^{\mathcal F_t}\). By
\cref{lem:screen-revelation}, the full-history value is
\[
U_t^{\mathcal F_t}(\mu)
=
\sup_{q,p}
\int_{\mathcal Y}\!\int_\Theta
\bigl[p(\theta,y)-c(y)q(\theta,y)\bigr]\,
\mu(d\theta)\nu_t(dy),
\]
where the supremum is over measurable full-history schedules satisfying
the pointwise IC--IR constraints. Fix any feasible full-history schedule
\((q,p)\). Its payoff is
\[
\int_{\mathcal Y}\!\int_\Theta
\bigl[p(\theta,y)-c(y)q(\theta,y)\bigr]\,
\mu(d\theta)\nu_t(dy),
\]
which is affine in \(\mu\). The feasibility of \((q,p)\) is determined
by the pointwise IC--IR constraints and is therefore independent of \(\mu\). Taking the supremum over all feasible schedules, we obtain that \(U_t^{\mathcal F_t}\) is convex.

Finally, the supremum of convex functions is convex. Therefore
\[
g(\mu)=\sup_t U_t^{\sigma(S_t)}(\mu),
\qquad
\widehat g(\mu)=\sup_t U_t^{\mathcal F_t}(\mu)
\]
are convex.
\end{proof}

\begin{lemma}[Extreme-point concavification]
\label{lem:screen-conc}
Let \(h:\Delta(\Theta)\to\mathbb R\) be bounded, convex, and upper semicontinuous. Then for every \(\mu\in\Delta(\Theta)\),
\[
(\operatorname{conc}h)(\mu)
=
\int_\Theta h(\delta_\theta)\,\mu(d\theta).
\]
\end{lemma}

\begin{proof}
For every \(\mu\in\Delta(\Theta)\), convexity of \(h\) gives
\[
h(\mu)
\le
\int_\Theta h(\delta_\theta)\,\mu(d\theta),
\]
because \(\mu\) is the barycenter of the degenerate posteriors \(\{\delta_\theta\}_{\theta\in\Theta}\). The right-hand side is affine in \(\mu\), hence concave, and it majorizes \(h\). Therefore
\[
(\operatorname{conc}h)(\mu)
\le
\int_\Theta h(\delta_\theta)\,\mu(d\theta).
\]

Conversely, since \(\operatorname{conc}h\) is concave and
\(\operatorname{conc}h\ge h\),
\[
(\operatorname{conc}h)(\mu)
\ge
\int_\Theta(\operatorname{conc}h)(\delta_\theta)\,\mu(d\theta)
\ge
\int_\Theta h(\delta_\theta)\,\mu(d\theta).
\]
Combining the two inequalities gives
\[
(\operatorname{conc}h)(\mu)
=
\int_\Theta h(\delta_\theta)\,\mu(d\theta).
\]

\end{proof}

\begin{lemma}[Screen-known-type]
\label{cor:screen-known-type}
Suppose \cref{ass:screen-regularity} holds. Then
\[
\bigl(\operatorname{conc}_f\widehat g\bigr)(S_0)
=
\int_\Theta \widehat g(\delta_\theta)\,S_0(d\theta),
\qquad
\bigl(\operatorname{conc}_f g\bigr)(S_0)
=
\int_\Theta g(\delta_\theta)\,S_0(d\theta).
\]
\end{lemma}

\begin{proof}
By \cref{lem:screen-convex}, \(g\) and \(\widehat g\) are convex. Since they are also bounded and upper semicontinuous, \cref{lem:screen-conc} applies to both functions and gives
\[
(\operatorname{conc}\widehat g)(S_0)
=
\int_\Theta \widehat g(\delta_\theta)\,S_0(d\theta),
\qquad
(\operatorname{conc}g)(S_0)
=
\int_\Theta g(\delta_\theta)\,S_0(d\theta).
\]
Since \(\operatorname{conc}_f g\) and
\(\operatorname{conc}_f\widehat g\) are upper semicontinuous at \(S_0\) by \cref{ass:screen-regularity}, \cref{lem:concf-conc} applies first to \(q=g\) and then to \(q=\widehat g\). Hence
\[
(\operatorname{conc}_f g)(S_0)
=
(\operatorname{conc}g)(S_0),
\qquad
(\operatorname{conc}_f\widehat g)(S_0)
=
(\operatorname{conc}\widehat g)(S_0).
\]
Combining the displayed equalities gives the result.
\end{proof}

\subsection{Proof of \cref{prop:value-representation}}
\begin{proof}
The lower bound follows from the finite date-0 experiment construction: for any finite Bayes-plausible distribution over posteriors, choose on each posterior branch an $\varepsilon$-optimal date-wise continuation selector. By \cref{lem:measurable-lifting} and \cref{lem:sequential-pasting}, these branchwise continuation equilibria can be pasted into a sequentially rational equilibrium of the induced calendar, attaining the corresponding weighted average up to $\varepsilon$. Taking the supremum over finite Bayes-plausible posterior distributions gives the lower bound.

The upper bound is exactly \cref{lem:upper-bound}. Combining the two inequalities gives the stated representation.
\end{proof}

\subsection{Proof of \cref{thm:nogain-delta-eps}}
\begin{proof}
Write
\[
g:=\sup_{t\in\mathbb T}U_t^{\sigma(S_t)},
\qquad
\widehat g:=\sup_{t\in\mathbb T}U_t^{\mathcal G_t}.
\]
By \cref{prop:value-representation}, the seller's optimal ex-ante value under the public state $\mathcal G_t$ equals $(\operatorname{conc}_f\widehat g)(S_0)$, and that under $\sigma(S_t)$ equals $(\operatorname{conc}_f g)(S_0)$. Hence $\mathcal G_t$ generates no additional value at the prior if and only if
\begin{equation}\tag{$\ast$}\label{eq:nogain-star}
(\operatorname{conc}_f\widehat g)(S_0)=(\operatorname{conc}_f g)(S_0).
\end{equation}
Since $\sigma(S_t)\subseteq\mathcal G_t$ gives $g\le\widehat g$ and
$\operatorname{conc}_f$ is monotone, the inequality
$(\operatorname{conc}_f g)(S_0)\le(\operatorname{conc}_f\widehat g)(S_0)$ always
holds; \eqref{eq:nogain-star} is therefore equivalent to its reverse.

Throughout we use the canonical identification of
$\mathrm{Aff}_c(\Delta(\Theta))$ with $C(\Theta)$. For $\lambda\in C(\Theta)$ the functional $\ell_\lambda(\mu):=\int_\Theta\lambda(\theta)\,\mu(d\theta)$ is affine and weakly continuous, so $\ell_\lambda\in\mathrm{Aff}_c(\Delta(\Theta))$; conversely, given $\ell\in\mathrm{Aff}_c(\Delta(\Theta))$ and setting $\lambda(\theta):=\ell(\delta_\theta)$, continuity of $\theta\mapsto\delta_\theta$ on the compact space $\Theta$ gives $\lambda\in C(\Theta)$, while affineness together with the weak density of finitely supported measures yields
$\ell=\ell_\lambda$. Writing $\ell_\varepsilon:=\ell_{\lambda_\varepsilon}$,
conditions \ref{cond:posterior-support}, \ref{cond:tightness}, and
\ref{cond:richer-cap} read, respectively,
\[
g\le\ell_\varepsilon\ \text{on }\Delta(\Theta),
\qquad
\ell_\varepsilon(S_0)\le(\operatorname{conc}_f g)(S_0)+\varepsilon,
\qquad
\widehat g\le\ell_\varepsilon\ \text{on }\Delta(\Theta).
\]

\medskip
\noindent\textit{Sufficiency.}\enspace
Fix $\varepsilon>0$ and let $\lambda_\varepsilon\in C(\Theta)$ satisfy
\ref{cond:posterior-support}--\ref{cond:richer-cap}; set $\ell_\varepsilon:=\ell_{\lambda_\varepsilon}$. By the richer-state cap, $\widehat g\le\ell_\varepsilon$ on $\Delta(\Theta)$. Since $\ell_\varepsilon$ is affine, every finite Bayes-plausible distribution $\sum_k\alpha_k\delta_{\mu_k}$ with barycenter $S_0$ satisfies
\[
\sum_k\alpha_k\widehat g(\mu_k)
\le\sum_k\alpha_k\ell_\varepsilon(\mu_k)
=\ell_\varepsilon(S_0),
\]
and taking the supremum over such distributions gives
$(\operatorname{conc}_f\widehat g)(S_0)\le\ell_\varepsilon(S_0)$. With the
tightness condition,
\[
(\operatorname{conc}_f\widehat g)(S_0)
\le\ell_\varepsilon(S_0)
=\int_\Theta\lambda_\varepsilon\,dS_0
\le(\operatorname{conc}_f g)(S_0)+\varepsilon .
\]
Letting $\varepsilon\downarrow0$ yields the reverse inequality; hence
\eqref{eq:nogain-star}.

\medskip
\noindent\textit{Necessity.}\enspace
Suppose \eqref{eq:nogain-star} holds, and set $\phi:=\operatorname{conc}_f\widehat g$. As a finite concavification, $\phi$ is concave and $\widehat g\le\phi$; by \cref{ass:regularity}, $\phi$ is upper semicontinuous at $S_0$. As in the proof of \cref{lem:concf-conc}, the supporting hyperplane theorem \citep[Section~5.12]{AliprantisBorder}, applied to the hypograph of $\phi$ at $(S_0,\phi(S_0)+\varepsilon)$, yields for each $\varepsilon>0$ an affine function $\ell_\varepsilon\in\mathrm{Aff}_c(\Delta(\Theta))$ with
\[
\widehat g\le\phi\le\ell_\varepsilon\ \text{on }\Delta(\Theta),
\qquad
\ell_\varepsilon(S_0)\le\phi(S_0)+\varepsilon .
\]
By \eqref{eq:nogain-star}, $\phi(S_0)=(\operatorname{conc}_f g)(S_0)$, so
$\ell_\varepsilon(S_0)\le(\operatorname{conc}_f g)(S_0)+\varepsilon$. Put
$\lambda_\varepsilon(\theta):=\ell_\varepsilon(\delta_\theta)$; by the
identification above $\lambda_\varepsilon\in C(\Theta)$ and
$\ell_\varepsilon(\mu)=\int_\Theta\lambda_\varepsilon\,d\mu$. The richer-state
cap holds because $U_t^{\mathcal G_t}(\mu)\le\widehat g(\mu)\le\int_\Theta
\lambda_\varepsilon\,d\mu$ for all $t\in\mathbb T$ and $\mu\in\Delta(\Theta)$;
posterior support then follows from $g\le\widehat g$; and tightness is the bound
$\int_\Theta\lambda_\varepsilon\,dS_0=\ell_\varepsilon(S_0)\le
(\operatorname{conc}_f g)(S_0)+\varepsilon$. As $\varepsilon>0$ was arbitrary,
this proves the claim.
\end{proof}

\subsection{Proof of \cref{prop:collapse-structure}}
\begin{proof}
Write
\[
g:=\sup_{t\in\mathbb T}U_t^{\sigma(S_t)},
\qquad
\widehat g:=\sup_{t\in\mathbb T}U_t^{\mathcal F_t}.
\]
By \cref{prop:value-representation}, the optimal ex-ante value of the
full-history calendar equals $(\operatorname{conc}_f\widehat g)(S_0)$. A terminal calendar is in particular a full-history calendar, so by \cref{def:collapse} the full-history calendar collapses to a terminal mechanism at $S_0$ if and only if the value attainable by terminal calendars equals $(\operatorname{conc}_f\widehat g)(S_0)$. It therefore suffices to compute that value.

By \cref{def:collapse}, a terminal calendar induces a finite Bayes-plausible distribution of posteriors and then, after each realized posterior $\mu$, applies a $\sigma(S_{t(\mu)})$-admissible selector at some date $t(\mu)\in\mathbb T$. Such a selector yields at most $U_{t(\mu)}^{\sigma(S_{t(\mu)})}(\mu)\le g(\mu)$, so for the inducing distribution $\sum_k\alpha_k\delta_{\mu_k}$, with barycenter $S_0$, the calendar's payoff is at most $\sum_k\alpha_k g(\mu_k)\le(\operatorname{conc}_f
g)(S_0)$. Hence every terminal calendar is bounded above by $(\operatorname{conc}_f g)(S_0)$.

Conversely, fix a finite Bayes-plausible distribution $\sum_{k=1}^K\alpha_k \delta_{\mu_k}$ with barycenter $S_0$, and for each $k$ choose a date $t_k$ and a $\sigma(S_{t_k})$-admissible selector at $\mu_k$ with value exceeding $g(\mu_k)-\varepsilon$. By the measurable-lifting and sequential-pasting argument (\cref{lem:measurable-lifting,lem:sequential-pasting}), the corresponding terminal calendar attains at least $\sum_{k=1}^K\alpha_k g(\mu_k)-\varepsilon$. Taking the supremum over all such distributions and letting $\varepsilon\downarrow0$, the value attainable by terminal calendars is exactly $(\operatorname{conc}_f
g)(S_0)$.

Collapse at $S_0$ is therefore equivalent to
\[
(\operatorname{conc}_f\widehat g)(S_0)=(\operatorname{conc}_f g)(S_0),
\]
i.e.\ to the full public state $\mathcal F_t=\sigma(S_t,Y_t)$ generating no additional value at the prior. By \cref{thm:nogain-delta-eps} applied with $\mathcal G_t=\mathcal F_t$, this holds if and only if, for every $\varepsilon>0$, there is a continuous $\lambda_\varepsilon\colon\Theta\to\mathbb R$ satisfying \cref{cond:posterior-support,cond:tightness} and \eqref{eq:terminal-support}, the last being \cref{cond:richer-cap} for $\mathcal G_t=\mathcal F_t$.
\end{proof}
\subsection{Proof of \cref{prop:diagnostic}}
\begin{proof}
Suppose the full-history calendar collapses to a terminal mechanism at $S_0$, and fix $\varepsilon>0$. By \cref{prop:collapse-structure}, there is a continuous $\lambda_\varepsilon\colon\Theta\to\mathbb R$ satisfying
\cref{cond:posterior-support,cond:tightness} and \eqref{eq:terminal-support}.
Taking the supremum over $t\in\mathbb T$ in \eqref{eq:terminal-support},
\[
\widehat g(\mu)\le\int_\Theta\lambda_\varepsilon(\theta)\,\mu(d\theta)
\qquad\forall\,\mu\in\Delta(\Theta).
\]
Since $\lambda_\varepsilon$ satisfies \cref{cond:posterior-support,cond:tightness},
it is among the prices admitted in the supremum defining $\mathcal N(S_0)$, whence
\[
\int_\Theta\lambda_\varepsilon(\theta)\,\mu(d\theta)
\le\sup_{\lambda}\int_\Theta\lambda(\theta)\,\mu(d\theta)
\qquad\forall\,\mu\in\Delta(\Theta).
\]
Combining the two displays gives $\widehat g(\mu)\le\sup_{\lambda}\int_\Theta
\lambda(\theta)\,\mu(d\theta)$ for every $\mu$; that is,
$(\mu,\widehat g(\mu))\notin\mathcal N(S_0)$ for all $\mu\in\Delta(\Theta)$.
\end{proof}

\section{Additional supplements}

\subsection{Primitive sufficient conditions for \cref{ass:regularity}}
\label{app:regularity-sufficient}

We record three sufficient conditions for \cref{ass:regularity}, in decreasing order of abstraction; the paper itself uses only the assumption. Write $\widehat g:=\sup_{t\in\mathbb T}U_t^{\mathcal G_t}$. Since $\Delta(\Theta)$ is compact and convex, the upper-semicontinuous concave envelope of a bounded-above $h$ admits the dual description
\[
(\operatorname{conc}h)(\mu)
=
\inf\{\ell(\mu):\ell\in\mathrm{Aff}_c(\Delta(\Theta)),\ \ell\ge h\},
\]
an infimum of continuous functions and hence upper semicontinuous; this is the envelope used in \cref{lem:concf-conc}. The conditions below either verify this agreement directly or give a local argument for upper semicontinuity of
$\operatorname{conc}_f\widehat g$ at $S_0$.

\paragraph{R1: continuation values.}
Suppose each $U_t^{\mathcal G_t}$ is upper semicontinuous and uniformly bounded above, and the date-wise supremum is locally finite at $S_0$, i.e.\ $\widehat g=\max_{t\in T_N}U_t^{\mathcal G_t}$ on a neighborhood $N$ of $S_0$ for some finite $T_N\subset\mathbb T$. Then $\widehat g$ is upper semicontinuous on $N$, and if in addition $\operatorname{conc}_f\widehat g=\operatorname{conc}\widehat g$ near $S_0$, then \cref{ass:regularity} holds. When $\Theta$ is finite this identity holds whenever $\widehat g$ is upper semicontinuous on the simplex, by finite-dimensional concavification; that, and local finiteness, are automatic when $\mathbb T$ is finite. Otherwise the identity is an additional requirement, since finite concavification may fall short of $\operatorname{conc}\widehat g$.

\paragraph{R2: game-theoretic primitives.}
The reduced-form regularity inputs in R1 hold, beyond \cref{ass:analytic-equilibrium}, if for each $t$: (a) $\mathcal H_t$ is compact and the admissible-mechanism and equilibrium correspondences are compact-valued and upper hemicontinuous; (b) $\pi_t(\omega,z,h,b)$ is jointly continuous and uniformly bounded above; and (c) $\mu\mapsto\mathbb P_t^\mu$ is weakly continuous. By (b) and (c), $\mu\mapsto\int_\Omega\pi_t\,d\mathbb P_t^\mu$ is continuous, so the maximum theorem~\citep[Theorem~17.31]{AliprantisBorder} renders each $U_t^{\mathcal G_t}$ upper semicontinuous. Continuity of $\pi_t$ in $\omega$ strengthens the measurability in \cref{ass:analytic-equilibrium} and is what carries the weakly continuous kernel through the integral.

\paragraph{R3: local condition at the prior.}
Suppose $S_0\in\operatorname{int}\Delta(\Theta)$, $\widehat g$ has a bounded upper-semicontinuous extension on a neighborhood $N$ of $S_0$, and for all $\mu$ in some neighborhood $M\subset N$ the value $(\operatorname{conc}_f\widehat g)(\mu)$ is approximated arbitrarily closely by finite splits supported in $N$. Then \cref{ass:regularity} holds, even if $\widehat g$ is irregular near $\partial\Delta(\Theta)$.

\medskip\noindent
R2 supplies the upper semicontinuity and boundedness inputs of R1; together with R1's local finite-date condition and finite-split adequacy, R1 implies \cref{ass:regularity}. R3 gives an alternative local sufficient condition. The applications of \cref{sec:Empirical-interpretation} have finite $\Theta$ and are covered by R1.

\subsection{Sufficient conditions for Selector collapse} \label{subsec:condi-selec-coll}
\begin{proposition}[Sufficient conditions for Selector collapse]
\label{thm:feasible-implementation}
Suppose that the continuation value collapse from \(\sigma(S_t,Y_t)\) to \(\mathcal G\). If at least one of \ref{item:FI1}--\ref{item:FI4} holds, then the selector collapse holds from \(\sigma(Z_t)\) to \(\mathcal G\).
\end{proposition}

\begin{enumerate}[label=\textup{(FI\arabic*)},leftmargin=*]

\item \label{item:FI1}
\textbf{Common optimal face.}
Suppose that, conditional on \((\mu,\mathcal G,Y_t)\), the induced
date-\(t\) continuation problem admits a feasible set
\[
\mathcal P_t(\mu,\mathcal G,Y_t)\subseteq\mathbb R^d
\]
and an objective
\[
p\longmapsto \langle c_t(\mu,\mathcal G),p\rangle,
\]
where the residual public state may affect feasibility but not the
objective coefficient. For each \((t,\mu,\mathcal G)\), there exists a
nonempty face
\[
F_t(\mu,\mathcal G)\subseteq\mathbb R^d
\]
such that, for every feasible realization of the residual public state,
\[
F_t(\mu,\mathcal G)
\subseteq
\argmax_{p\in \mathcal P_t(\mu,\mathcal G,Y_t)}
\langle c_t(\mu,\mathcal G),p\rangle,
\]
and there exists a measurable selector
\[
p_t^*(\mu,\mathcal G)\in F_t(\mu,\mathcal G)
\]
that is feasible for every such realization.

\item \label{item:FI2}
\textbf{Common dual certificate.}
Suppose that, conditional on \((\mu,\mathcal G,Y_t)\), the induced
date-\(t\) continuation problem can be written as
\[
\max_{x\in \mathcal X_t(\mu,\mathcal G,Y_t)}
\ \langle c_t(\mu,\mathcal G),x\rangle,
\qquad
\mathcal X_t(\mu,\mathcal G,Y_t)
=
\{x:A_t(\mu,\mathcal G,Y_t)x\le b_t(\mu,\mathcal G,Y_t)\}.
\]
For each \((t,\mu,\mathcal G)\), there exist a dual vector
\[
\lambda_t^*(\mu,\mathcal G)\ge 0,
\]
an active index set
\[
I_t^*(\mu,\mathcal G),
\]
and a measurable selector
\[
x_t^*(\mu,\mathcal G)
\]
such that, for every feasible realization of the residual public state:

\begin{enumerate}[label=\textup{(\alph*)},leftmargin=*]
\item \(\lambda_t^*(\mu,\mathcal G)\) is dual-feasible and dual-optimal;
\item \(x_t^*(\mu,\mathcal G)\in \mathcal X_t(\mu,\mathcal G,Y_t)\);
\item complementary slackness holds with the same active set
\(I_t^*(\mu,\mathcal G)\);
\item the primal and dual objective values coincide at
\[
\big(x_t^*(\mu,\mathcal G),\lambda_t^*(\mu,\mathcal G)\big).
\]
\end{enumerate}

\item \label{item:FI3}
\textbf{Unique \(\mathcal G\)-optimal continuation mechanism.}
For every posterior \(\mu\in\Delta(\Theta)\), there exists a unique
continuation mechanism--payoff-selection pair
\[
(\Gamma_t^*(\mu),\rho_t^*(\mu))
\]
such that \(\Gamma_t^*(\mu)\) is \(\mathcal G\)-measurable,
\[
\int_\Omega \rho_t^*(\omega)\,\mathbb P_t^\mu(d\omega)
=
U_t^{\mathcal G}(\mu),
\]
\[
\rho_t^*\in \Sel\big(\mathcal R_t^{\Gamma_t^*(\mu)}\big),
\]
and this pair is admissible under the full public history
\(\mathcal F_t\).

\item \label{item:FI4}
\textbf{Threshold invariance.}
There exists a \(\mathcal G\)-measurable family of candidate continuation
rules
\[
\{\Gamma_t^*(\mu)\}_{\mu\in\Delta(\Theta)}
\]
such that, for each \(\mu\), some payoff selection
\[
\rho_t^*(\mu)\in \Sel\big(\mathcal R_t^{\Gamma_t^*(\mu)}\big)
\]
satisfies
\[
\int_\Omega \rho_t^*(\omega)\,\mathbb P_t^\mu(d\omega)
=
U_t^{\mathcal G}(\mu),
\]
and for every posterior \(\mu\) and every refinement of the public history
inside the same \(\mathcal G\)-event, the ranking of candidate continuation
rules is unchanged.

\end{enumerate}

\begin{proof}
Fix \(\mu\in\Delta(\Theta)\). By value equality, it suffices to construct a
\(\mathcal G\)-measurable continuation rule \(\Gamma_t\) and a payoff
selection
\[
\rho\in\Sel\big(\mathcal R_t^{\Gamma_t}\big)
\]
such that
\[
\int_\Omega \rho(\omega)\,\mathbb P_t^\mu(d\omega)
=
U_t^{\mathcal F_t}(\mu).
\]

Under \ref{item:FI1}, the common-face selector is feasible for every refinement of the residual public state and lies in a face consisting of full-history optimizers. Hence it is optimal under the full public history.

Under \ref{item:FI2}, the common primal--dual certificate verifies optimality of the same \(\mathcal G\)-measurable choice for every refinement of the residual public state.

Under \ref{item:FI3}, the unique continuation mechanism attaining \(U_t^{\mathcal G}(\mu)\) is admissible under the full public history and, by value equality, also attains \(U_t^{\mathcal F_t}(\mu)\).

Under \ref{item:FI4}, the same continuation rule remains optimal after every refinement of the public history inside a given \(\mathcal G\)-event.

In each case, there exists a \(\mathcal G\)-measurable continuation rule that is optimal under the full public history. Thus \(\mathcal O_t^{\mathcal F_t}\) admits a \(\mathcal G\)-measurable selector.
\end{proof}

\subsection{Iterated representation of the continuation value} \label{sub:Iterated representation}
For a finite horizon \(T\), the continuation value unfolds as
\[
\begin{aligned}
U_t^{\mathcal G_t}(\mu)
=
\max\Bigg\{
&V_t^{\mathcal G_t}(\mu),\\
&\int_\Omega
\max\Bigg[
V_{t+1}^{\mathcal G_{t+1}}\bigl(S_{t+1}(\omega_{t+1})\bigr),\\
&\qquad
\int_\Omega
\max\Bigg[
V_{t+2}^{\mathcal G_{t+2}}\bigl(S_{t+2}(\omega_{t+2})\bigr),
\ldots,\\
&\qquad\quad
\int_\Omega
V_T^{\mathcal G_T}\bigl(S_T(\omega_T)\bigr)\,
\mathbb P_{T-1}^{S_{T-1}(\omega_{T-1})}(d\omega_T)
\Bigg]\,
,...,
\mathbb P_{t+1}^{S_{t+1}(\omega_{t+1})}(d\omega_{t+2})
\Bigg]\,
\mathbb P_t^\mu(d\omega_{t+1})
\Bigg\}.
\end{aligned}
\]
The iterated form expresses \(U_t^{\mathcal G_t}(\mu)\) entirely through the static truncation values \(\{V_s^{\mathcal G_s}\}_{s\ge t}\), but it does so by repeated integration over the posterior process: each terminal value \(V_s^{\mathcal G_s}\) is evaluated at the realized posterior \(S_s\),  compared with the value of deferring one more date. 

\end{document}